\documentclass[11pt,letterpaper]{article}
\usepackage[T1]{fontenc}
\usepackage[utf8]{inputenc}
\usepackage{lmodern}
\usepackage[margin=1in]{geometry}
\usepackage{microtype}
\usepackage{amsmath,amssymb}
\usepackage{graphicx}
\usepackage{float} % Keep main-text figures with their relevant passages.
\usepackage{booktabs,longtable,array,calc,multirow}
\usepackage{caption}
\usepackage{pdflscape}
\usepackage[numbers,sort&compress,square]{natbib}
\usepackage{xurl}
\usepackage{hyperref}
\hypersetup{colorlinks=true,linkcolor=black,citecolor=black,urlcolor=black,pdftitle={Sure Enough to Share? Overconfidence in GenAI Video Discernment is Associated with Sharing Intentions}}
\usepackage{bookmark}
\title{Sure Enough to Share? Overconfidence in GenAI Video Discernment is Associated with Sharing Intentions}
\author{%
  Haoning Xue\textsuperscript{1},
  Sijia Qian\textsuperscript{2},
  Christopher Calabrese\textsuperscript{3},\\
  Cuihua Shen\textsuperscript{4},
  Laramie Taylor\textsuperscript{4},
  Yixuan Wang\textsuperscript{1}\\[0.6em]
  {\small \textsuperscript{1}University of Utah}\\
  {\small \textsuperscript{2}University of North Carolina at Charlotte}\\
  {\small \textsuperscript{3}Clemson University}\\
  {\small \textsuperscript{4}University of California, Davis}
}
\date{}
\begin{document}
\maketitle
\begin{abstract}
Generative Artificial Intelligence (GenAI) is blurring the line between synthesized and recorded videos, but less is known about people's awareness of their discernment abilities and the consequences of miscalibration. This study examines three forms of overconfidence in GenAI video discernment (i.e., overestimation, overplacement, and overprecision) and their associations with source judgments and sharing intentions in an online study (N = 973). Participants evaluated videos depicting everyday scenes and achieved 62.5\% accuracy. Three forms of overconfidence showed distinct patterns: participants overplaced themselves by 12.8 percentile points, but overprecision showed consistent positive associations with sharing intentions and the tendency to classify videos as human-made. Age, GenAI usage, and AI literacy showed consistent positive associations with all three forms of overconfidence. These findings highlight the importance of distinguishing the three forms of overconfidence to better understand GenAI engagement and incorporating metacognition into interventions to support informed media consumption.
\end{abstract}

\section{Introduction}\label{introduction}

From deceptive deepfakes \citep{Kobis2021_3017} to seemingly innocent wildlife videos \citep{Feltrin2026_3203}, generative Artificial Intelligence (GenAI) is blurring the line between synthesized and recorded videos, challenging the seeing-is-believing assumption. This concern extends beyond deceptive deepfakes \citep{Kobis2021_3017} and obvious AI slop \citep{Curtis2026_2986} to AI-generated videos of ordinary scenes such as cats playing, glaciers melting, or people cooking. These videos resemble the everyday footage people scroll past online and may not prompt them to consider whether they are AI-generated. Mistaking these scenes for authentic footage may carry fewer immediate consequences than misjudgments about elections or public emergencies, but these errors can still shape perceptions of reality, and sharing these GenAI videos as authentic may pass those distorted perceptions on to others.

This concern involves not only whether people can detect GenAI videos but also whether they recognize the limits of their abilities. Empirical studies suggest that people discern GenAI images and videos from recorded ones only slightly better than chance and tend to judge videos as human-made \citep{Cooke2025_2965,Davodi2026_2969,Sarfraz2025_3018}. For example, participants achieved about 60\% accuracy overall but only 43\% for videos generated using text-to-video models \citep{Sarfraz2025_3018}. Yet people do not always accurately assess their GenAI video discernment \citep{Chen2025_3062,Fatima2026_3012}. This discrepancy between perceived and actual performance reflects miscalibration as overconfidence or underconfidence. Of particular concern is overconfidence, which occurs when individuals assess their performance or abilities more favorably or with greater certainty than warranted \citep{Moore2008_1270}. Miscalibration can carry consequences beyond the judgment itself. Overconfidence is associated with poorer performance and reduced information-seeking across contexts such as finance \citep{Daniel2015_3013}, medicine \citep{Saposnik2016_3016}, and education \citep{Dunlosky2012_1207}. In digital information contexts, overconfident individuals have been shown to be more susceptible to misinformation and more likely to share it \citep{Lyons2021_1135,Skurka2026_2999}. However, less is known about to what extent people are aware of their GenAI video discernment and the consequences when they are miscalibrated.

Overconfidence takes three distinct forms: overestimation, overplacement, and overprecision \citep{Moore2008_1270}. Research has mostly focused on overestimation, the belief that one's ability is better than it actually is. Overplacement is the exaggerated belief that one is better than others, the social aspect of overconfidence. Overprecision is excessive certainty that a specific judgment is correct or ``being too sure you know the truth'' \citep[p. 1]{Moore2017_1231}. Specifically, these forms separate general expectations about GenAI video discernment accuracy (e.g., ``I can tell GenAI videos from real ones''), social comparisons (e.g., ``I'm better than most at detecting GenAI''), and certainty about one case (e.g., ``I'm certain this video is AI-generated''). Though related, these forms do not necessarily occur together; their prevalence and relationships vary across tasks and conditions, and they are linked to different outcomes \citep{Grezo2021_3000,Moore2008_1270}. Thus, these forms of overconfidence may also differ in how common they are and in how people engage with GenAI videos. A more systematic examination is needed to examine how miscalibration varies across three forms of overconfidence.

The present study examines how accurately people assess their GenAI video discernment ability and how (mis)calibration relates to their willingness to share these videos. Using videos depicting everyday health, environmental, and entertainment scenes, this study makes two contributions. First, we move beyond discernment accuracy to metacognition, providing a fuller account of people's authenticity judgments in relation to their intentions to share GenAI videos. Second, we distinguish three forms of overconfidence by comparing their calibration patterns and distinct associations with GenAI video discernment and sharing intentions. Together, the findings can inform more targeted interventions in improving GenAI video consumption.

\subsection{GenAI Video Discernment and Metacognitive Calibration}\label{genai-video-discernment-and-metacognitive-calibration}

People can distinguish AI-generated videos from human-made ones slightly above chance: GenAI discernment varies across modality \citep{Cooke2025_2965}, video length \citep{Davodi2026_2969}, and generation methods (e.g., text-to-video vs. image-to-video; \citep{Sarfraz2025_3018}). Yet people are not always aware of their actual performance. Research shows both excessive certainty in individual deepfake judgments and overestimation of overall GenAI discernment abilities \citep{Chen2025_3062,Kobis2021_3017}. Individual differences in calibration also exist. One study found that 22.4\% of participants were overconfident in their discernment abilities, and 27.6\% were underconfident \citep{Fatima2026_3012}. These studies provide initial insight into miscalibration in GenAI discernment, but a more systematic examination is needed.

Distinguishing these three forms is important because they involve different judgment processes and reflect different aspects of self-assessment \citep{Moore2017_1231}. Overestimation reflects imperfect knowledge of one's ability and task difficulty \citep{Moore2008_1270}. Overplacement often stems from imperfect knowledge of others and egocentric self-focus in social comparison \citep{Moore2017_1231}. Overprecision can arise from \emph{error neglect}, when people fail to recognize alternative explanations or reflect on potential ways their judgments might be wrong \citep{Moore2023_3040}.

The relationships among these three forms vary in both direction and strength. Moore and Healy \citep{Moore2008_1270} found that while overprecision persisted across task difficulty, participants overestimated their performance but underplaced themselves on difficult quizzes and underestimated their performance but overplaced themselves on easy quizzes. Overestimation and overplacement were moderately correlated in cancer misinformation discernment (r = .53; \citep{Lyons2026_3065}). Across the three forms, only overestimation and overplacement were highly correlated in financial literacy (r = .817; \citep{Voros2021_3064}), but overplacement and overprecision were strongly correlated in interval estimation tasks (r = .60; \citep{Duttle2016_3066}). Three forms of overconfidence do not have to co-occur or be equally pronounced. Thus, we examine whether and to what extent people's self-assessments of estimated performance, relative standing, and judgment certainty align with their actual performance in GenAI video discernment.

\begin{quote}
\textbf{RQ1.} How calibrated are people about their GenAI video discernment in terms of overestimation, overplacement, and overprecision?
\end{quote}

Beyond distinguishing AI-generated from human-made videos, discernment also involves the bias toward one source over the other. Signal detection theory distinguishes these aspects as sensitivity and response bias, respectively \citep{Batailler2022_2599}. The deception-detection literature defines truth bias as the tendency to judge information as truthful \citep{Levine2014_564}. A related tendency in GenAI discernment concerns perceived origin, as people tend to classify AI-generated content as authentic or human-made \citep{Kobis2021_3017,Pehlivanoglu2026_3060}.

This tendency may be associated with overconfidence because they may share underlying cognitive processes. For example, while error neglect relates to overprecision because people fail to recognize potential ways their judgments might be wrong, it can also be reflected in overlooking the possibility that a video could be created by different sources, thereby reinforcing existing response bias. Studies on deception detection found a positive association between truth bias and confidence \citep{DePaulo1997_3090}. However, it is still unclear whether this relationship extends to overconfidence in GenAI video discernment, and what specific form of overconfidence it involves. Examining this association can pinpoint which form is most relevant to this tendency to lean toward a particular source.

\begin{quote}
\textbf{RQ2.} How does overconfidence in GenAI video discernment (i.e., overestimation, overplacement, overprecision) relate to response tendency?
\end{quote}

\subsection{Overconfidence and Sharing Intentions}\label{overconfidence-and-sharing-intentions}

Distinguishing three forms of overconfidence is also important for understanding the intention to share GenAI videos. Sharing intentions capture the willingness to circulate content without necessarily recognizing its source, particularly when video sources are not clearly disclosed or accurately labeled. However, AI-generated videos circulating as real events may distort perceptions of what exists, what is possible, and what has occurred.

Existing evidence linking overconfidence to information sharing has largely focused on overplacement. Overplacement has been related to the intention to engage with misinformation \citep{Lyons2021_1135}, especially among people with the News Finds Me perception \citep{Skurka2026_2999}. One possible explanation concerns the social functions of sharing. People who overplaced themselves were more likely to share information to display competence cues \citep{Anderson2012_3067}. Information sharing can allow people to showcase their knowledge and satisfy similar impression-management goals \citep{Berger2014_3095}. Thus, overplacement may encourage sharing as an opportunity to demonstrate competence and gain social recognition. Perceived capability offers another explanation. Perceived self-efficacy has been positively associated with knowledge sharing in virtual communities and on social media \citep{Hsu2007_3079,Kim2020_3077}. Thus, overestimation may relate to stronger sharing intentions, as people who feel capable of detecting GenAI may be more comfortable sharing it, regardless of actual performance.

Overprecision may also relate to sharing intentions because of excessive certainty. Research on attitude certainty suggests that people who feel highly certain about their judgments are more likely to advocate for them \citep{Cheatham2017_3080}. Similarly, people with excessive certainty about a video's source may be more willing to share their judgments or engage with the video. However, sharing can also serve information-seeking purposes, such that people who are uncertain may share to solicit advice \citep{Berger2014_3095}. While these findings suggest different pathways through which overconfidence may relate to sharing intentions, it remains unclear how the three forms of overconfidence relate to sharing. Therefore, we propose the following research question.

\begin{quote}
\textbf{RQ3.} How does overconfidence in GenAI video discernment (i.e., overestimation, overplacement, overprecision) relate to sharing intentions?
\end{quote}

\subsection{Individual Differences in Overconfidence}\label{individual-differences-in-overconfidence}

Empirical evidence on individual differences in overconfidence is mixed. A meta-analysis found a small and nonsignificant gender difference across 39 studies \citep{Bandiera2022_3096}. Age-related differences vary across forms, but only overprecision was positively related to age \citep{Prims2017_3047}. Political ideology also matters. Conservatives showed more judgment confidence than liberals \citep{Ruisch2021_3104}, and political extremity on both sides was associated with greater overconfidence \citep{Prooijen2019_3113}. Lastly, individual domain-relevant experience can also shape self-assessments. Self-reported AI literacy was associated with stronger overestimation \citep{Fernandes2026_3108}. Similarly, beginners' initial learning experiences can boost overconfidence \citep{Sanchez2018_1208}. Since overconfidence varies across individual characteristics and experiences, we examine how these factors relate to three forms of overconfidence.

\begin{quote}
\textbf{RQ4.} How are individual characteristics associated with overconfidence in GenAI video discernment?
\end{quote}

\section{Methods}\label{methods}

\subsection{Study Design and Procedure}\label{study-design-and-procedure}

We conducted a 2 (video source: AI-generated vs. human-made) \ensuremath{\times} 3 (context: health, environment, and entertainment) within-subjects experiment programmed in Qualtrics. All study procedures were approved by the university's Institutional Review Board. We collected \emph{n} = 1058 complete responses from U.S. adults recruited through CloudResearch Connect in November 2025. Recruitment used quotas to approximate U.S. population distribution for age, gender, and race. Further, we enabled the Qualtrics fraud detection feature to identify potential bot and duplicate responses \citep{Qualtrics2026_3020}. We excluded responses from participants who reported video-loading issues (\emph{n} = 34), those with reCAPTCHA scores below 0.5 (\emph{n} = 50), and one duplicated response (\emph{n} = 1). The final sample included \emph{N} = 973 participants. See Table~\ref{tab:s1} for a summary of demographics.

Each participant viewed 18 videos drawn from a pool of 42 experimental stimuli (see below for stimuli development). All videos were presented without audio to maintain consistency across stimuli generated with and without sound. This pool included 21 pairs of AI-generated and human-made videos, matched by topic, with seven topics in each context. An automated sampling script randomly selected three AI-generated and three human-made videos, covering six distinct topics within each context. This was meant to minimize potential fatigue from repeated exposure to the same topic. Each participant therefore viewed 18 distinct topics in random order, with nine AI-generated and nine human-made videos, and never saw both versions of the same topic.

After providing informed consent, participants reported their demographics and completed pre-task self-assessments of their GenAI video discernment. They estimated (a) how many of the 18 videos they expected to identify correctly and (b) their percentile standing among 100 American adults. Participants also reported GenAI usage, AI literacy, digital skill literacy, and sensation-seeking tendency. During each trial, the video played on a continuous loop, and playback controls were disabled. Participants classified the video as AI-generated or human-made, rated their certainty in that judgment, and indicated how likely they were to share it. After completing all 18 trials, participants again estimated their number of correct judgments in this study and their percentile rank. Although not the primary aim of this study, these assessments were intended to capture potential changes in self-evaluation. Participants then received a debriefing showing their total number of correct judgments and the videos they had classified correctly and incorrectly. See Figure~\ref{fig:s1} for an illustration of study procedures. The median completion time was 13 minutes.

\subsection{Stimuli Development}\label{stimuli-development}

We developed the experimental stimuli in two steps: (a) creating AI-generated videos and identifying human-made counterparts and (b) pilot testing the videos to select the final stimulus set. First, we used OpenAI's Sora 2, Google's Veo 3, and Midjourney to generate videos on a wide range of everyday health, environmental, and entertainment topics. We followed a consistent prompt structure specifying the depicted scene and visual presentation, then screened the videos for topic relevance and visual realism. We selected human-made counterparts for each AI-generated video from publicly available footage to match the topic and depicted scenes.

Second, we pilot-tested 58 candidate videos with \emph{n} = 100 participants recruited on CloudResearch Connect. Each participant evaluated 12 random videos, classified their sources, and rated perceived message sensation value \citep{Palmgreen2002_1970} and perceived eeriness \citep{Ho2017_3029,Wu2021_3027} on a seven-point scale (Tables~\ref{tab:s2}--\ref{tab:s3}). We used Rasch modeling estimates of item difficulty and model fit to inform video selection \citep{Robitzsch2025_3031}. The final set includes 42 videos (i.e., 21 topic pairs of AI-generated and human-made videos), with seven topic pairs in each context. See Table~\ref{tab:s4} for summary statistics and Supplemental Methods for detailed development procedures, selection criteria, and pilot results.

\subsection{Measures}\label{measures}

\subsubsection{GenAI Video Discernment Performance and Response Tendency}\label{genai-video-discernment-performance-and-response-tendency}

\emph{Classification accuracy}. We coded trial-level correctness as correct (1) or incorrect (0). Individual-level classification accuracy (\emph{M} = 0.62, \emph{SD} = 0.14, \emph{Min} = 0.22, \emph{Max} = 1.00) was calculated as the proportion of videos classified correctly across the 18 trials for each participant.

\emph{Discernment sensitivity and response tendency}. We calculated signal detection sensitivity (d\ensuremath{\prime}) and response criterion (\emph{c}) to distinguish two dimensions of GenAI video discernment: the ability to differentiate between sources and the tendency to favor one source response over the other \citep{Stanislaw1999_3055}. Sensitivity (\emph{M} = 0.66, \emph{SD} = 0.75) reflects one's ability to distinguish AI-generated from human-made videos, with higher values indicating better performance in GenAI video discernment. Response criterion (\emph{M} = 0.024, \emph{SD} = 0.412) reflects source response tendency: positive values indicate a tendency toward human-made responses, negative values indicate a tendency toward AI-generated responses, and zero indicates a neutral response criterion. We calculated these measures from hit (i.e., correctly classified AI-generated videos) and false-alarm (i.e., human-made videos misclassified as AI-generated) rates, treating AI-generated videos as the signal. We applied a loglinear correction before calculating hit and false-alarm rates to avoid infinite values. See Table~\ref{tab:1} for a summary table.

\subsubsection{Estimated Confidence}\label{estimated-confidence}

\emph{Estimated proportion correct}. We measured pre-task performance confidence with one question: ``In this study, you will watch 18 short videos and decide whether each video was generated by AI or a human. Before you begin, how many do you think you will correctly identify?'' We divided their estimated number correct by 18 to obtain the estimated proportion correct (\emph{M} = 0.62, \emph{SD} = 0.19, \emph{Min} = 0.11, \emph{Max} = 1.00).

\emph{Estimated percentile rank}. We measured pre-task estimated percentile rank (\emph{M} = 62.82, \emph{SD} = 20.80, \emph{Min} = 3, \emph{Max} = 100) with one question: ``If we lined up 100 American adults based on how good they are at spotting AI-generated videos, where would you place yourself?''

\emph{Estimated certainty}. For each trial, we measured per-trial certainty (\emph{M} = 63.71\%, \emph{SD} = 23.56\%, \emph{Min} = 0\%, \emph{Max} = 100\%) with one question: ``On a scale from 0\% to 100\%, how confident are you about your judgment?'' These ratings were rescaled to 0-1 for analysis and averaged across the 18 trials to obtain individual-level estimated certainty (\emph{M} = 0.64, \emph{SD} = 0.17).

\subsubsection{Overconfidence}\label{overconfidence}

We operationalized three forms of overconfidence: estimating one's ability, exaggerating one's ability relative to others, and having excessive certainty in one's judgment \citep{Moore2017_1231}. Each measure compared estimated confidence with a corresponding actual performance benchmark. Although we use the conventional term overconfidence, the resulting scores are signed with positive values indicating overconfidence, negative values indicating underconfidence, and zero indicating no obvious discrepancy between estimated confidence and actual performance. See Table~\ref{tab:1} for descriptive statistics of estimated confidence and overconfidence and Table~\ref{tab:s5} for correlations among three forms of overconfidence.

\emph{Overestimation}. We calculated overestimation (\emph{M} = -0.006, \emph{SD} = 0.221) as the estimated proportion correct minus the actual proportion correct \citep{Moore2008_1270}.

\emph{Overplacement}. We calculated overplacement by comparing the estimated percentile with their actual performance percentile within the study sample. To calculate actual performance percentiles, we ranked participants by their classification accuracy, assigned average ranks to ties \citep{Hudgens2002_3037}, and converted these midranks to percentiles to avoid raw extremes using 100 \ensuremath{\times} (midrank - 0.5)/N (N is the number of participants; \citep{Hyndman1996_3052}). We then calculated individual-level overplacement (\emph{M} = 12.82, \emph{SD} = 32.81) as the estimated percentile minus the actual percentile \citep{Li2025_3039}.

\emph{Overprecision}. For overprecision, we assessed certainty in every video judgment using direct confidence ratings \citep{Li2025_3039}. Based on participants' estimated certainty in every video judgment, we calculated trial-level overprecision (\emph{M} = 0.012, \emph{SD} = 0.518) as estimated certainty minus classification correctness per trial. We averaged the trial-level scores to calculate the individual-level overprecision (\emph{M} = 0.012, \emph{SD} = 0.205). This item-confidence approach directly captures estimated certainty relative to actual performance, with positive scores indicating excessive certainty. However, a known limitation is that this approach cannot cleanly distinguish excessive certainty from overestimation of performance \citep{Prims2017_3047}.

\subsubsection{Other Measures}\label{other-measures}

\emph{GenAI usage.} We assessed participants' use of 10 popular GenAI tools \citep{Zhu2024_2336} with one question: ``How often do you interact with the following generative AI tools?'' on a five-point scale. We averaged ratings across tools to summarize GenAI usage frequency (\emph{M} = 1.43, \emph{SD} = 0.49) (See Table~\ref{tab:s6} for the full list).

\emph{AI literacy}. We assessed AI literacy (\emph{M} = 5.05, \emph{SD} = 1.00, Cronbach's \ensuremath{\alpha} = .89) with three items focusing on GenAI detection on a seven-point scale \citep{Carolus2023_3033} (Table~\ref{tab:s7}).

\emph{Digital skill literacy}. We included digital skill literacy (\emph{M} = 5.14, \emph{SD} = 1.11, Cronbach's \ensuremath{\alpha} = .88) using four items from the Digital Media and Information Literacy Scale \citep{Qian2026_3034}, which captures perceived competence in using digital technologies (Table~\ref{tab:s8}).

\emph{Sensation seeking}. We included sensation seeking (\emph{M} = 2.48, \emph{SD} = 0.84, Cronbach's \ensuremath{\alpha} = .82) using eight questions on a five-point scale to capture individual tendency to seek stimulation that may shape responses to GenAI videos \citep{Hoyle2002_1680} (Table~\ref{tab:s9}).

\emph{Sharing intentions}. We measured the intention to share a video (\emph{M} = 2.44, \emph{SD} = 1.79) at each trial with one question on a seven-point scale: ``If you saw this video on your social media feed, how likely would you be to share it in a chat with your family or friends?'' This choice was informed by research showing that interpersonal sharing emphasizes information usefulness to and strong social connection with the recipient \citep{Barasch2014_3022}.

\subsection{Analytical Strategy}\label{analytical-strategy}

To answer RQ1 about calibration in GenAI video discernment, we summarized actual performance, estimated confidence, and the three overconfidence measures. We used one-sample \emph{t}-tests to examine whether individual-level overconfidence scores differed from zero.

To answer RQ2 about the associations between overconfidence and human-source response tendency, we fitted linear regressions predicting response criterion. We first included three forms of overconfidence as independent variables, then adjusted for six socio-demographic variables (age, gender, race, education, income, political affiliation) and four individual characteristics (GenAI usage, AI literacy, digital skill literacy, and sensation seeking).

To answer RQ3 about the associations between overconfidence and sharing intentions, we first fitted similar individual-level linear regressions predicting sharing intentions. We conducted supplementary trial-level analyses of sharing intentions using linear mixed-effects models with the \emph{lmerTest} package in R \citep{Kuznetsova2017_99}. These models included trial-level overprecision and individual-level overestimation and overplacement, although these were measured at the individual level. In addition to random intercepts for participants and trial order, these models also accounted for video context, actual source, perceived source, and the same socio-demographics and individual characteristics.

To answer RQ4 about the individual characteristics associated with overconfidence, we fitted separate individual-level linear regressions for overestimation, overplacement, and overprecision. Predictors included the six socio-demographic and four individual characteristics described above. We adjusted p-values for multiple comparisons using the Benjamini-Hochberg procedure across the three overconfidence models.

\section{Results}\label{results}

\subsection{Calibration in GenAI Video Discernment}\label{calibration-in-genai-video-discernment}

\begingroup
\small
\setlength{\tabcolsep}{4pt}
\renewcommand{\arraystretch}{1.12}
\captionof{table}{Descriptive statistics of GenAI video discernment performance, response tendency, estimated confidence, and overconfidence at the individual and trial levels.}\label{tab:1}
{ % do not increment counter
\begin{longtable*}[]{@{}
  >{\raggedright\arraybackslash}p{(\linewidth - 10\tabcolsep) * \real{0.4076}}
  >{\raggedright\arraybackslash}p{(\linewidth - 10\tabcolsep) * \real{0.1185}}
  >{\raggedright\arraybackslash}p{(\linewidth - 10\tabcolsep) * \real{0.1185}}
  >{\raggedright\arraybackslash}p{(\linewidth - 10\tabcolsep) * \real{0.1185}}
  >{\raggedright\arraybackslash}p{(\linewidth - 10\tabcolsep) * \real{0.1185}}
  >{\raggedright\arraybackslash}p{(\linewidth - 10\tabcolsep) * \real{0.1185}}@{}}
\toprule\noalign{}
\begin{minipage}[b]{\linewidth}\raggedright
\textbf{Measure}
\end{minipage} & \begin{minipage}[b]{\linewidth}\centering
\textbf{M}
\end{minipage} & \begin{minipage}[b]{\linewidth}\centering
\textbf{SD}
\end{minipage} & \begin{minipage}[b]{\linewidth}\centering
\textbf{Min}
\end{minipage} & \begin{minipage}[b]{\linewidth}\centering
\textbf{Max}
\end{minipage} & \begin{minipage}[b]{\linewidth}\centering
\textbf{Median}
\end{minipage} \\
\midrule\noalign{}
\endhead
\bottomrule\noalign{}
\endlastfoot
\multicolumn{6}{@{}>{\raggedright\arraybackslash}p{(\linewidth - 10\tabcolsep) * \real{0.9999} + 10\tabcolsep}@{}}{%
\textbf{Individual level (\emph{N} = 973)}} \\
Classification accuracy (\%) & 62.48 & 13.69 & 22.22 & 100 & 61.11 \\
Sensitivity (d\ensuremath{\prime}) & 0.66 & 0.75 & -1.52 & 3.29 & 0.55 \\
Response criterion & 0.02 & 0.41 & -1.64 & 1.64 & 0.00 \\
Estimated accuracy (\%) & 61.93 & 19.29 & 11.11 & 100 & 61.11 \\
Estimated percentile & 62.82 & 20.80 & 3.00 & 100 & 65 \\
Actual percentile & 50.00 & 28.68 & 0.05 & 99.90 & 46.09 \\
Mean judgment certainty (\%) & 63.71 & 17.07 & 1.00 & 100 & 65.56 \\
Overestimation (percentage points) & -0.55 & 22.14 & -66.67 & 55.56 & 0.00 \\
Overplacement (percentile points) & 12.82 & 32.81 & -75.35 & 89.93 & 13.80 \\
Overprecision (percentage points) & 1.23 & 20.52 & -73.72 & 60.94 & 1.78 \\
\multicolumn{6}{@{}>{\raggedright\arraybackslash}p{(\linewidth - 10\tabcolsep) * \real{0.9999} + 10\tabcolsep}@{}}{%
\textbf{Trial level (\emph{n} = 17,514)}} \\
Classification correctness (0-1) & 0.62 & 0.48 & 0 & 1 & 1 \\
Judgment certainty (\%) & 63.71 & 23.56 & 0 & 100 & 65 \\
Overprecision (percentage points) & 1.23 & 51.80 & -100 & 100 & -12.00 \\
\end{longtable*}
}
\endgroup

\begin{figure}[H]
\centering
\includegraphics[width=\linewidth,height=.74\textheight,keepaspectratio]{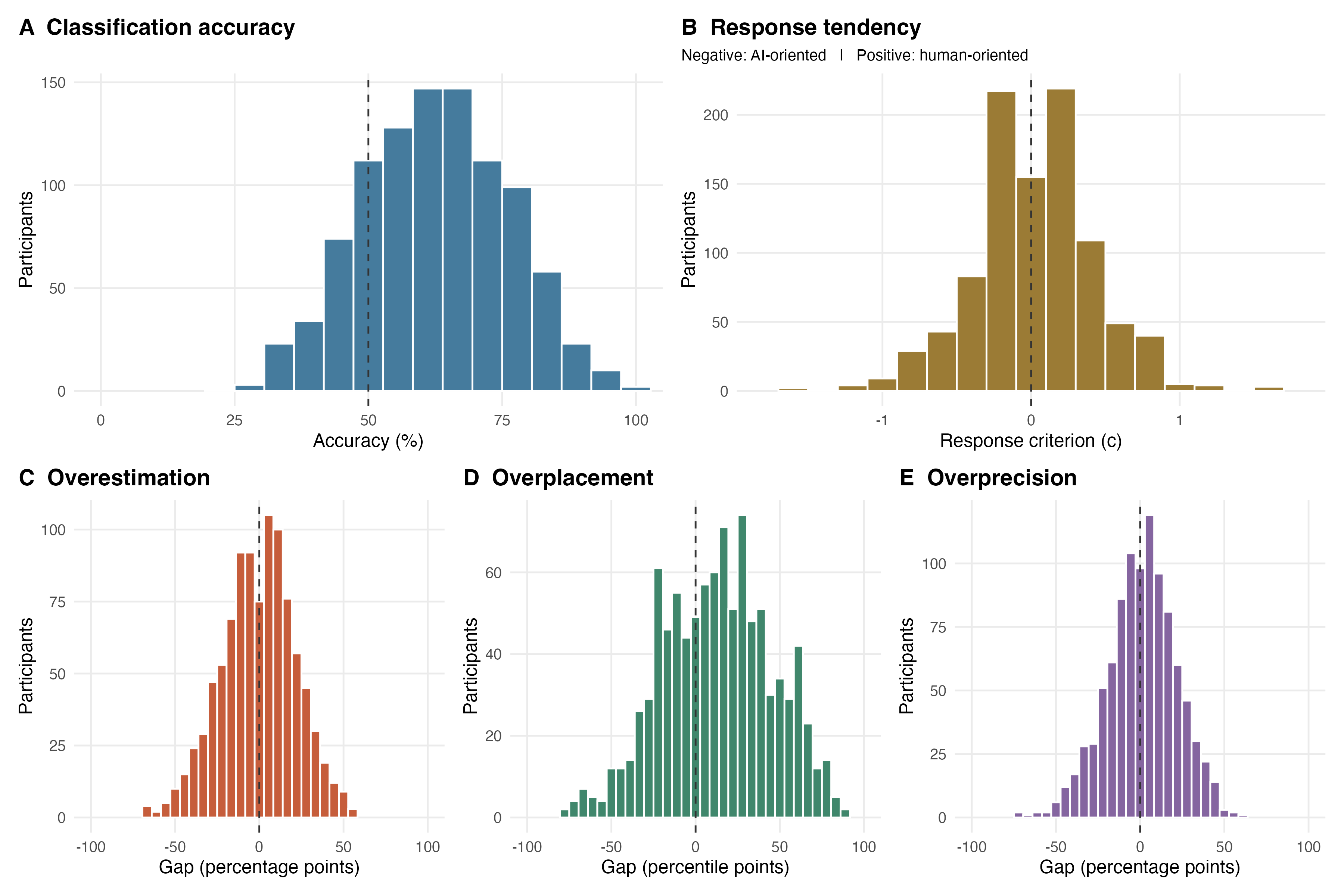}
\caption{Distributions of (A) classification accuracy, (B) response tendency, (C) overestimation, (D) overplacement, and (E) overprecision. Dashed vertical lines mark chance accuracy (50\%) in Panel A and zero response criterion or calibration gap in Panels B-E. Positive response criterion indicates a tendency to classify videos as human-made; negative criterion indicates a tendency to classify them as AI-generated. Positive calibration gaps indicate overconfidence, and negative gaps indicate underconfidence.}\label{fig:1}
\end{figure}

Participants in this study were able to distinguish AI-generated from human-made videos with an average accuracy of 62.48\% (\emph{SD} = 13.69\%), exceeding the 50\% chance level (\emph{t} = 28.42, \emph{p} \textless{} .001) (see Figure~\ref{fig:1} Panel A for accuracy distribution). Participants were slightly more accurate in classifying human-made videos (63.45\%) than AI-generated videos (61.51\%; \emph{t} = 2.04, \emph{p} = .042). GenAI video discernment ability was also reflected in a positive signal detection sensitivity score (\emph{M} = 0.66, \emph{SD} = 0.75). Participants showed no significant tendency toward either AI or human classifications with a close-to-zero average response criterion (\emph{M} = 0.024, \emph{SD} = 0.41) (Figure~\ref{fig:1} Panel B).

Participants exhibited significant overplacement, ranking themselves on average 12.82 percentile points (\emph{SD} = 32.81) above their actual position in the sample (\emph{t} = 12.19, \emph{p} \textless{} .001). However, neither overestimation, the gap between estimated and actual accuracy (\emph{M} = -0.006, \emph{SD} = 0.221; \emph{t} = -0.77, \emph{p} = .440), nor overprecision, the gap between mean judgment certainty and actual accuracy (\emph{M} = 0.012, \emph{SD} = 0.205; \emph{t} = 1.87, \emph{p} = .062) differed significantly from zero. See Figure~\ref{fig:1} Panels C-E for distribution. 

\begin{figure}[H]
\centering
\includegraphics[width=\linewidth,height=.74\textheight,keepaspectratio]{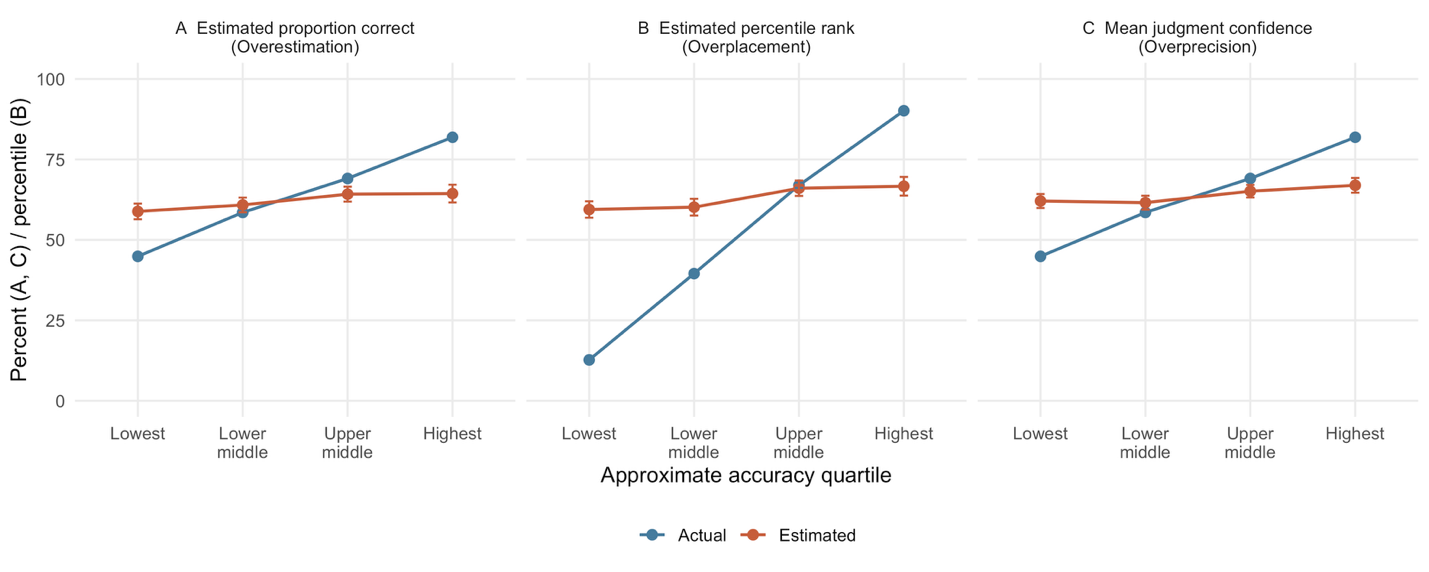}
\caption{Estimated confidence and actual performance across approximate accuracy quartiles for (A) overestimation, (B) overplacement, and (C) overprecision. Orange shows estimated accuracy, estimated percentile standing, and mean item-level certainty, respectively; blue shows the corresponding actual performance benchmarks. Error bars represent two-sided, t-based 95\% confidence intervals for the group means. See Table~\ref{tab:s11} for means and standard deviations.}\label{fig:2}
\end{figure}

These near-zero averages did not mean participants were well calibrated. We examined calibration across performance levels by grouping participants into four quartiles (Figure~\ref{fig:2}; Table~\ref{tab:s11}), finding that the lowest-performing participants showed the greatest overconfidence on average, a pattern consistent with the Dunning-Kruger effect \citep{Kruger1999_27}. Specifically, this group achieved 45\% accuracy and ranked the 13th percentile, while reporting an expected proportion correct of 59\%, an estimated percentile rank of 59\%, and an estimated certainty of 62\%. In contrast, the top group achieved 82\% accuracy and ranked the 90th percentile, but reported an expected proportion correct of 64\%, an estimated percentile of 67\%, and a mean estimated certainty of 67\%. Thus, participants' estimated confidence varied less than their actual performance: underconfidence among higher performers and overconfidence among lower performers offset one another, contributing to the near-zero averages in overestimation and overprecision.

For RQ2, we fitted individual-level models to examine the relationship between overconfidence and response tendency, with positive values indicating a response tendency toward human classifications (see Figure~\ref{fig:3} Panel A and Table~\ref{tab:s12} for details). With or without adjustment for covariates, overprecision was positively associated with response tendency. This suggests that participants with excessive certainty in their judgments were more inclined to classify a video as human-made.

\subsection{\texorpdfstring{Overprecision is Associated with Sharing Intentions }{Overprecision is Associated with Sharing Intentions }}\label{overprecision-is-associated-with-sharing-intentions}

\begin{figure}[H]
\centering
\includegraphics[width=\linewidth,height=.74\textheight,keepaspectratio]{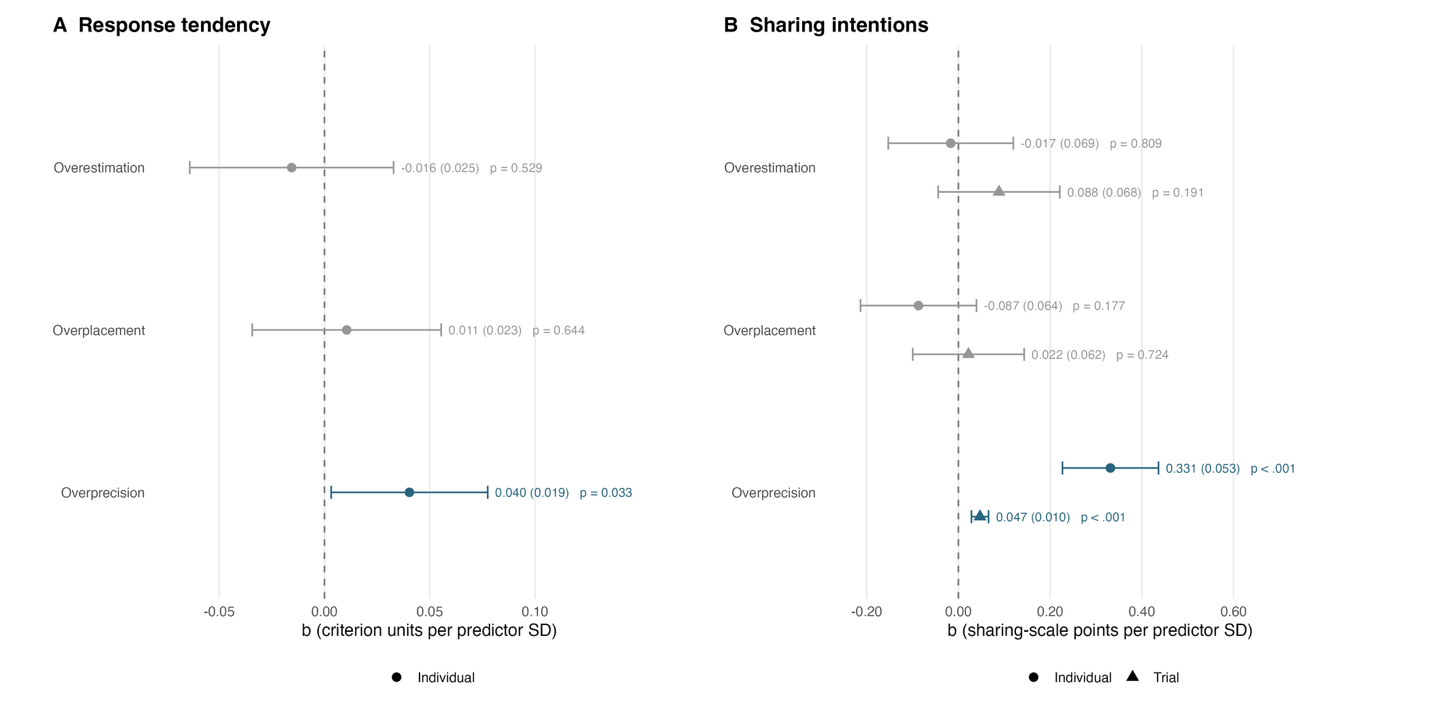}
\caption{Adjusted associations of overconfidence with (A) individual-level response tendency and (B) sharing intentions at the individual and trial levels. Blue estimates have unadjusted p \textless{} .05; gray estimates have p \ensuremath{\geq} .05. The dashed vertical line marks zero. Annotations report b (SE) and p values. Numeric predictors are standardized; response criterion and sharing intentions retain their original scales. See Tables~\ref{tab:s12}--\ref{tab:s14} for full models.}\label{fig:3}
\end{figure}

RQ3 examined the association between overconfidence and sharing intentions. Overprecision showed the most consistent positive association with sharing intentions across analyses. In the joint models including all three forms, overprecision was positively associated with sharing intentions at both the individual (\emph{b} = 0.33, \emph{p} \textless{} .001) and trial (\emph{b} = 0.05, \emph{p} \textless{} .001) levels (see Figure~\ref{fig:3} Panel B and Tables~\ref{tab:s13}--\ref{tab:s14} for details). Overestimation and overplacement were not significantly related to sharing intentions at either level in adjusted models.

We also fitted separate models that included only one form at a time to examine how these results depend on adjustment for other forms of overconfidence (see Tables~\ref{tab:s15}--\ref{tab:s16} for model details). Across both individual and trial levels, all three forms were positively associated with sharing intentions. As all three forms are correlated with each other (see Table~\ref{tab:s5} for correlations), the single-form models described each form's association with sharing intentions without accounting for the variance shared across all three forms. The relationships with overestimation and overplacement were sensitive to that distinction and became nonsignificant after accounting for other forms of overconfidence. On the contrary, the positive relationship between overprecision and sharing intentions was consistent at both individual and trial levels.

\begin{figure}[H]
\centering
\includegraphics[width=\linewidth,height=.74\textheight,keepaspectratio]{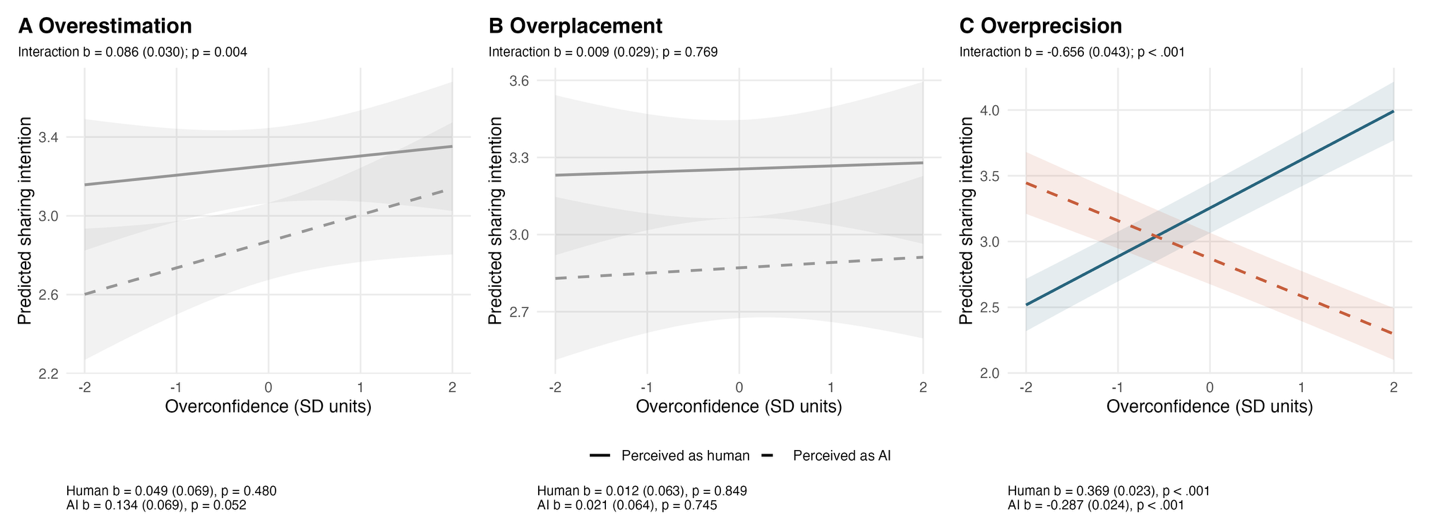}
\caption{Predicted sharing intentions by (A) overestimation, (B) overplacement, and (C) overprecision, moderated by videos perceived as human-made or AI-generated. Solid lines represent perceived human-made videos, and dashed lines represent perceived AI-generated videos. Significant simple slopes (p \textless{} .05) are blue for perceived human-made videos and orange for perceived AI-generated videos; nonsignificant slopes are grayed out. Shaded bands are model-based 95\% confidence intervals for the predicted mean sharing intention.}\label{fig:4}
\end{figure}

We further investigated whether this positive relationship varied across perceived source, that is, whether participants classified a video as AI-generated or human-made, independent of its actual source (see Figure~\ref{fig:4} and Table~\ref{tab:s14} for details). Perceived source significantly moderated the trial-level associations for overprecision and overestimation. Overprecision showed opposite associations across perceived sources: it was positively associated with sharing intentions for videos perceived as human-made but negatively associated with sharing intentions for videos perceived as AI-generated. However, overestimation was positively associated with intentions for videos perceived as AI-generated, but this association was not statistically significant (\emph{b} = 0.13, \emph{p} = .052). It was also not significantly associated with intentions for videos perceived as human-made (\emph{b} = 0.05, \emph{p} = .480). As participants made judgments on perceived source and sharing intentions in the same trial, these results remain correlational.

\subsection{Individual Differences in Overconfidence}\label{individual-differences-in-overconfidence-1}

\begin{figure}[H]
\centering
\includegraphics[width=\linewidth,height=.74\textheight,keepaspectratio]{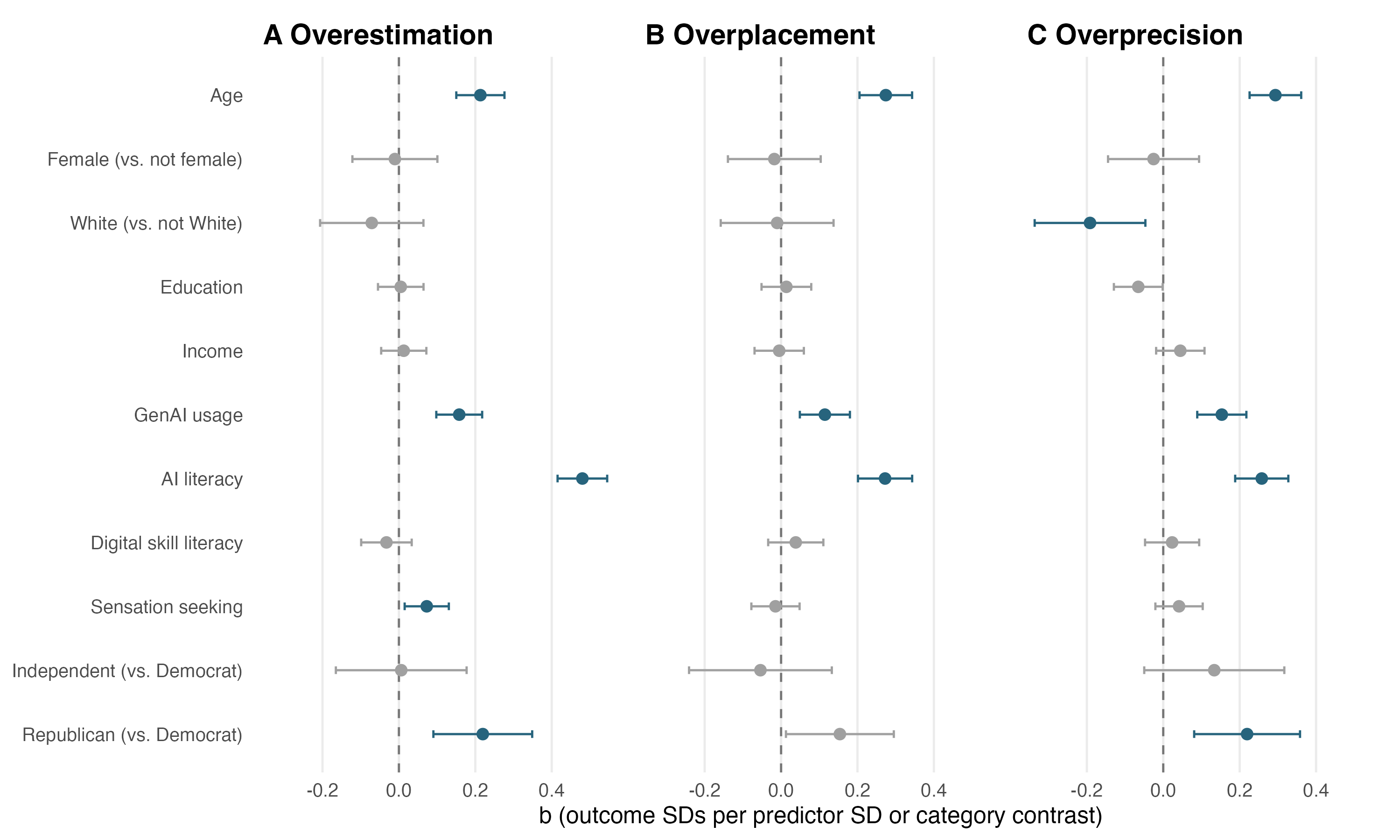}
\caption{Individual characteristics associated with (A) overestimation, (B) overplacement, and (C) overprecision. Points show regression coefficients (b) from separate individual-level linear regressions for the three standardized overconfidence outcomes; numeric predictors are also standardized. Significant estimates after adjustment for multiple comparisons were marked in blue; nonsignificant estimates were grayed out. The dashed vertical line marks zero. See Table~\ref{tab:s17} for full models.}\label{fig:5}
\end{figure}

RQ4 examined the individual characteristics associated with overconfidence. Across three individual-level regression models, age, GenAI usage, and AI literacy were consistently significant predictors of overconfidence with adjustment for multiple comparisons (Figure~\ref{fig:5}; Table~\ref{tab:s17}). Specifically, older participants (\emph{b} = 0.21-0.29, all \emph{q} \textless{} .001) and those reporting higher levels of GenAI usage (\emph{b} = 0.11-0.16, all \emph{q} \textless{} .01) and AI literacy (\emph{b} = 0.26-0.48, all \emph{q} \textless{} .001) were more likely to be overconfident across three forms. Further, compared to Democrat affiliation, Republican affiliation was associated with higher levels of overestimation (\emph{b} = 0.22, \emph{q} = .003) and overprecision (\emph{b} = 0.22, \emph{q} = .006), but the difference in overplacement (\emph{b} = 0.15, \emph{q} = .077) did not survive adjustment. Figure~\ref{fig:s2} further compares empirical performance and self-assessments across these individual characteristics and shows different patterns. Older adults showed lower performance but not proportionately lower self-assessment. Participants reporting higher GenAI usage and AI literacy showed better performance but inflated self-assessment. Republicans had lower performance despite comparable self-assessment.

\section{Discussion}\label{discussion}

This study systematically examined overconfidence in GenAI video discernment. While participants distinguished AI-generated videos slightly above chance with no obvious response bias, three findings stood out. First, participants overplaced themselves by 12.8 percentile points, but the minimal average overestimation or overprecision masked notable individual variation. Second, despite the prominence of overplacement, overprecision showed consistent positive associations with sharing intentions and the tendency to classify videos as human-made. Lastly, age, GenAI usage, and AI literacy showed consistent positive associations with all three forms of overconfidence. These findings show the importance of distinguishing the three forms of overconfidence, as not all are equally relevant to source judgments and sharing intentions.

First, the prominence of overplacement highlights the distinction between evaluating one's performance and comparing it with others' performance. This distinction may reflect the fact that people generally have less information about others than about themselves \citep{Moore2008_1270}. People are therefore more likely to overrate their standing relative to others than to overestimate their own performance. A similar pattern emerged in cancer misinformation discernment, where participants showed moderate overplacement but underestimated their discernment abilities \citep{Lyons2026_3065}. However, the near-zero average overestimation and overprecision do not necessarily suggest accurate individual calibration. Overconfidence among low-performing individuals and underconfidence among high-performing individuals may offset one another. This pattern aligns with evidence of both overconfidence and underconfidence in AI-manipulated video detection \citep{Fatima2026_3012}. Overall, these findings suggest that overconfidence varied across both overconfidence forms and individuals, underscoring the importance of examining metacognition beyond a single form or group averages.

Second, despite the prominence of overplacement, overprecision was associated with a greater human response tendency. These tendencies may share a common basis in whether an initial judgment receives further scrutiny. Human response tendency parallels truth bias: people generally assume honesty unless something raises suspicion \citep{Levine2014_564}. Although video authenticity is not the same as truthfulness, the seeing-is-believing assumption implies an analogous starting point for GenAI video discernment, as viewers might intuitively accept visuals as a record of actual events. Reconsidering this assumption demands careful consideration, which does not happen automatically. Metacognition may help activate that reconsideration. When an initial human-source judgment comes to mind and feels right, this feeling of rightness may reduce the need for further scrutiny or reconsideration of alternatives \citep{Thompson2011_3115}. This lack of careful consideration is also key to the error-neglect origin of overprecision, such that overlooking potential errors can lead to excessive certainty in people's judgments \citep{Moore2023_3040}. This interpretation is consistent with deception-detection research where confidence has long been linked to truth bias \citep{DePaulo1997_3090}. Therefore, the small positive association observed here might indicate a common process of insufficient scrutiny, but this correlational evidence does not establish that causal mechanism.

Among three forms of overconfidence, only overprecision showed a consistent positive association with sharing intentions at both individual and trial levels. Since overprecision and sharing intentions were both measured on each trial, this association could reflect shared measurement context. However, this association remained significant after aggregation at the individual level, suggesting it extends beyond judgments about individual videos, although this does not rule out measurement-level influence. In addition, all three forms were positively associated with sharing intentions when examined separately, broadly consistent with previous research linking overplacement to misinformation engagement \citep{Lyons2021_1135}. Previous research suggests different pathways through which overconfidence may encourage sharing, but this finding suggests these associations may not be independent, and previous findings on individual forms may partly reflect their overlap with other forms. It highlights the importance of accounting for shared variance to distinguish each form's contribution.

This association between overprecision and sharing intentions varied across perceived source: this association was positive for videos perceived as human-made but negative for those perceived as AI-generated. This contrasting pattern may partly reflect how people perceive the value attributed to each source. Research on AI disclosure penalty shows that content labeled as AI-generated is associated with lower perceived effort and is consistently devalued \citep{Lim2026_3122,Raj2026_3118}. When a video is perceived as human-made, viewers may evaluate it using conventional criteria, in which excessive certainty in their source judgment may amplify that intention and reduce hesitation to share content they find worthwhile. However, perceiving a video as AI-generated may diminish its perceived value, and that excessive certainty about its origin may reinforce that reluctance to share. Therefore, overprecision may strengthen viewers' willingness to act on their source judgments rather than uniformly being associated with higher sharing intentions. Sharing intentions should also be distinguished from the willingness to circulate synthetic content as authentic. Viewers may knowingly share AI-generated videos for entertainment or discussion. Assessing whether and how viewers communicate a video's perceived source would help clarify when such sharing may risk misleading others.

Lastly, age, GenAI usage, and AI literacy showed consistent positive associations with three forms of overconfidence in GenAI video discernment. First, this age association extends previous findings on overprecision \citep{Prims2017_3047} to other forms. However, previous research suggests that older adults underestimated algorithmic knowledge \citep{Chung2026_3125}, but age was unrelated to overconfidence in misinformation discernment \citep{Lyons2021_1135}. Further, the associations with self-reported GenAI usage and AI literacy align with evidence linking self-reported AI understanding with overestimation \citep{Fernandes2026_3108}. We also identify distinct patterns in performance-assessment comparisons that could inform targeted interventions. For example, participants reporting higher GenAI usage and AI literacy performed better, but their inflated self-assessments suggest that interventions targeting these individuals should prioritize confidence calibration rather than discernment skills alone. Older adults' lower performance, without a proportionate decline in self-assessments, suggests that both discernment and confidence calibration need support. Given the mixed evidence, the study's findings need replication to determine whether they generalize across discernment contexts and provide a reliable basis for targeted interventions.

This study highlights the value of incorporating metacognition into GenAI video discernment research and interventions. First, it is important to distinguish among forms of overconfidence and account for individual variation because the most pronounced form may not be the most behaviorally relevant, and group averages may mask individual miscalibrations. Second, by examining accuracy, response tendency, and metacognition together, this study provides a more comprehensive approach to understanding GenAI discernment. This approach also suggests that intervention and training should consider confidence calibration alongside detection accuracy. Overprecision's consistent associations with source judgments and sharing intentions make it a promising focus for future intervention research.

\subsection{Limitations and Future Research}\label{limitations-and-future-research}

This study has three major limitations. First, the study's findings partly depend on how we measured the three forms of overconfidence. Although we used established approaches, different measures of the same form do not always converge and may correlate only weakly \citep{Li2025_3039}. The measures also differed in timing and unit of analysis: overestimation and overplacement used pre-task measurement, but overprecision was measured for each trial and then aggregated at the individual level. Although this approach captured certainty in specific judgments, we cannot rule out the possibility that overprecision's associations partly reflect how it was measured. Future studies should use alternative measures to examine whether these findings hold across operationalizations.

Second, presenting videos without audio in a structured viewing task limits the findings' generalizability to everyday settings. Online content is often more complex, combining AI-generated material with human editing. Therefore, the study's findings may not precisely reflect people's actual discernment and confidence calibration. Future research should investigate how people respond to actual GenAI content in more realistic settings.

Lastly, this study's findings are correlational and do not establish causal relationships, especially since source judgments, estimated certainty, and sharing intentions were measured within each trial. Asking participants to judge the source may make perceived source influence their sharing decisions more than it would during everyday viewing.

Beyond these limitations, this study's findings suggest that metacognition is a promising focus for future interventions to support GenAI video discernment and informed sharing decisions. Building on previous research that performance feedback alone can reduce confidence without improving discernment accuracy \citep{Lyons2021_1135}, future research could explore whether feedback on overconfidence or metacognitive prompts can improve discernment and calibration and influence sharing decisions. These approaches could complement existing digital literacy interventions by helping people recognize the limits of their abilities.

\clearpage
\bibliographystyle{unsrtnat}
\bibliography{references}

@article{Kobis2021_3017,
  author = {K\"{o}bis, Nils C. and Dole\v{z}alov\'{a}, Barbora and Soraperra, Ivan},
  title = {Fooled Twice: {People} Cannot Detect Deepfakes but Think They Can},
  journal = {iScience},
  volume = {24},
  number = {11},
  pages = {103364},
  year = {2021},
  doi = {10.1016/j.isci.2021.103364}
}

@article{Feltrin2026_3203,
  author = {Feltrin, Andrea Natan},
  publisher = {The Ecological Citizen},
  title = {{When images outnumber encounters: Synthetic wilderness and ecological presence in an age of AI}},
  journal = {The Ecological Citizen},
  volume = {9},
  number = {2},
  pages = {168-176},
  year = {2026},
}

@misc{Curtis2026_2986,
  author = {Curtis, Liam},
  title = {The {TikTok} {AI} {Slop} {Report}},
  year = {2026},
  month = {jun},
  url = {https://www.kapwing.com/resources/the-tiktok-ai-slop-report/}
}

@article{Cooke2025_2965,
  author = {Cooke, Di and Edwards, Abigail and Barkoff, Sophia and Kelly, Kathryn},
  publisher = {Association for Computing Machinery},
  title = {As {Good} as a {Coin} {Toss:} {Human} {Detection} of {AI-Generated} {Content}},
  journal = {Commun. ACM},
  volume = {68},
  number = {10},
  pages = {100--109},
  year = {2025},
  address = {New York, NY, USA},
  doi = {10.1145/3729417}
}

@inproceedings{Davodi2026_2969,
  author = {Davodi, Catherine E. and Barrington, Sarah and Farid, Hany and Cooper, Emily A.},
  title = {Perceptual {Judgments} of {Video} {Authenticity:} {An} {Examination} of {Viewing} {Duration,} {Confidence,} {Content,} and {Strategies}},
  booktitle = {Proceedings of the IEEE/CVF Conference on Computer Vision and Pattern Recognition},
  pages = {10679-10686},
  year = {2026}
}

@article{Sarfraz2025_3018,
  author = {Sarfraz, Danyal and Karim, Raja and Kim, KwanMyung},
  title = {Synthetic {Realities:} {Evaluating} {Human} {Ability} to {Distinguish} {AI-Generated} {Videos} from {Real} {Footage}},
  journal = {IASDR Conference Series},
  year = {2025},
  url = {https://dl.designresearchsociety.org/iasdr/iasdr2025/pictorials/18}
}

@article{Chen2025_3062,
  author = {Chen, Chen and Goh, Dion Hoe-Lian},
  publisher = {John Wiley \& Sons, Ltd},
  title = {The {Role} of {Self-Efficacy,} {Critical} {Thinking,} and {Media} {Literacy} in {Human} {Deepfake} {Video} {Detection}},
  journal = {Proceedings of the Association for Information Science and Technology},
  volume = {62},
  number = {1},
  pages = {897-902},
  year = {2025},
  doi = {10.1002/pra2.1310}
}

@inproceedings{Fatima2026_3012,
  author = {Fatima, Kaniz and Wu, Y. Kelly and Uzun, Ersin},
  publisher = {Association for Computing Machinery},
  title = {Beyond the {Naked} {Eye:} {Empirical} {Study} of {How} {People} {Perceive,} {Detect,} and {Respond} to {AI-Manipulated} {Videos}},
  booktitle = {Proceedings of the 2026 CHI Conference on Human Factors in Computing Systems},
  series = {CHI '26},
  year = {2026},
  address = {New York, NY, USA},
  doi = {10.1145/3772318.3791799}
}

@article{Moore2008_1270,
  author = {Moore, Don A. and Healy, Paul J.},
  publisher = {American Psychological Association},
  title = {The Trouble with Overconfidence.},
  journal = {Psychological Review},
  volume = {115},
  pages = {502-517},
  year = {2008},
  doi = {10.1037/0033-295X.115.2.502}
}

@article{Daniel2015_3013,
  author = {Daniel, Kent and Hirshleifer, David},
  title = {Overconfident {Investors,} {Predictable} {Returns,} and {Excessive} {Trading}},
  journal = {Journal of Economic Perspectives},
  volume = {29},
  number = {4},
  pages = {61--88},
  year = {2015},
  doi = {10.1257/jep.29.4.61}
}

@article{Saposnik2016_3016,
  author = {Saposnik, Gustavo and Redelmeier, Donald and Ruff, Christian C. and Tobler, Philippe N.},
  title = {Cognitive Biases Associated with Medical Decisions: A Systematic Review},
  journal = {BMC Medical Informatics and Decision Making},
  volume = {16},
  number = {1},
  pages = {138},
  year = {2016},
  doi = {10.1186/s12911-016-0377-1}
}

@article{Dunlosky2012_1207,
  author = {Dunlosky, John and Rawson, Katherine A.},
  title = {Overconfidence Produces Underachievement: {Inaccurate} Self Evaluations Undermine Students' Learning and Retention},
  journal = {Improving Self-Monitoring and Self-Regulation of Learning: From Cognitive Psychology to the Classroom},
  volume = {22},
  number = {4},
  pages = {271-280},
  year = {2012},
  doi = {10.1016/j.learninstruc.2011.08.003}
}

@article{Lyons2021_1135,
  author = {Lyons, Benjamin A. and Montgomery, Jacob M. and Guess, Andrew M. and Nyhan, Brendan and Reifler, Jason},
  publisher = {Proceedings of the National Academy of Sciences},
  title = {Overconfidence in News Judgments Is Associated with False News Susceptibility},
  journal = {Proceedings of the National Academy of Sciences},
  volume = {118},
  number = {23},
  pages = {e2019527118},
  year = {2021},
  doi = {10.1073/pnas.2019527118}
}

@article{Skurka2026_2999,
  author = {Skurka, Chris and Cheng, Zicheng and Goyanes, Manuel and Gil de Z\'{u}\~{n}iga, Homero},
  title = {News {Finds} {Me} as the Illusion of Competence: Evidence for Overconfidence in Discernment of Political Misinformation},
  journal = {Human Communication Research},
  volume = {52},
  number = {1},
  pages = {11-23},
  year = {2026},
  doi = {10.1093/hcr/hqaf015}
}

@article{Moore2017_1231,
  author = {Moore, Don A. and Schatz, Derek},
  publisher = {John Wiley \& Sons, Ltd},
  title = {The Three Faces of Overconfidence},
  journal = {Social and Personality Psychology Compass},
  volume = {11},
  number = {8},
  pages = {e12331},
  year = {2017},
  doi = {10.1111/spc3.12331}
}

@article{Grezo2021_3000,
  author = {Gre\v{z}o, Mat\'{u}\v{s}},
  publisher = {Emerald Group Publishing Limited},
  title = {{Overconfidence and financial decision-making: a meta-analysis}},
  journal = {Review of Behavioral Finance},
  volume = {13},
  number = {3},
  pages = {276-296},
  year = {2021},
  address = {Leeds},
  doi = {10.1108/RBF-01-2020-0020}
}

@article{Moore2023_3040,
  author = {Moore, Don A.},
  publisher = {American Psychological Association},
  title = {Overprecision Is a Property of Thinking Systems.},
  journal = {Psychological Review},
  volume = {130},
  number = {5},
  pages = {1339-1350},
  year = {2023},
  address = {US},
  doi = {10.1037/rev0000370}
}

@article{Lyons2026_3065,
  author = {Lyons, Benjamin A. and King, Andy J and Kaphingst, Kimberly A},
  title = {Overconfidence in Ability to Discern Cancer Misinformation: A Conceptual Replication and Extension},
  journal = {Human Communication Research},
  volume = {52},
  number = {1},
  pages = {66-75},
  year = {2026},
  doi = {10.1093/hcr/hqaf017}
}

@article{Voros2021_3064,
  author = {V\"{o}r\"{o}s, Zs\'{o}fia and Szab\'{o}, Zolt\'{a}n and Kehl, D\'{a}niel and Kov\'{a}cs, Oliv\'{e}r B\'{e}la and Papp, Tam\'{a}s and Schepp, Zolt\'{a}n},
  publisher = {John Wiley \& Sons, Ltd},
  title = {The Forms of Financial Literacy Overconfidence and Their Role in Financial Well-Being},
  journal = {International Journal of Consumer Studies},
  volume = {45},
  number = {6},
  pages = {1292-1308},
  year = {2021},
  doi = {10.1111/ijcs.12734}
}

@article{Duttle2016_3066,
  author = {Duttle, Kai},
  publisher = {John Wiley \& Sons, Ltd},
  title = {COGNITIVE {SKILLS} {AND} {CONFIDENCE:} {INTERRELATIONS} {WITH} {OVERESTIMATION,} {OVERPLACEMENT} {AND} {OVERPRECISION}},
  journal = {Bulletin of Economic Research},
  volume = {68},
  number = {S1},
  pages = {42-55},
  year = {2016},
  doi = {10.1111/boer.12069}
}

@article{Batailler2022_2599,
  author = {Batailler, C\'{e}dric and Brannon, Skylar M. and Teas, Paul E. and Gawronski, Bertram},
  publisher = {SAGE Publications Inc},
  title = {A {Signal} {Detection} {Approach} to {Understanding} the {Identification} of {Fake} {News}},
  journal = {Perspectives on Psychological Science},
  volume = {17},
  number = {1},
  pages = {78-98},
  year = {2022},
  doi = {10.1177/1745691620986135}
}

@article{Levine2014_564,
  author = {Levine, Timothy R.},
  publisher = {SAGE Publications Inc},
  title = {Truth-Default {Theory} {(TDT):} {A} {Theory} of {Human} {Deception} and {Deception} {Detection}},
  journal = {Journal of Language and Social Psychology},
  volume = {33},
  number = {4},
  pages = {378-392},
  year = {2014},
  doi = {10.1177/0261927X14535916}
}

@article{Pehlivanoglu2026_3060,
  author = {Pehlivanoglu, Didem and Zhu, Mengdi and Zhen, Jialong and Gagnon-Roberge, Aude A. and Kern, Rebecca K. and Woodard, Damon and Cahill, Brian S. and Ebner, Natalie C.},
  title = {Is This Real? {Susceptibility} to Deepfakes in Machines and Humans},
  journal = {Cognitive Research: Principles and Implications},
  volume = {11},
  number = {1},
  pages = {3},
  year = {2026},
  doi = {10.1186/s41235-025-00700-y}
}

@article{DePaulo1997_3090,
  author = {DePaulo, Bella M. and Charlton, Kelly and Cooper, Harris and Lindsay, James J. and Muhlenbruck, Laura},
  publisher = {SAGE Publications Inc},
  title = {The {Accuracy-Confidence} {Correlation} in the {Detection} of {Deception}},
  journal = {Personality and Social Psychology Review},
  volume = {1},
  number = {4},
  pages = {346-357},
  year = {1997},
  doi = {10.1207/s15327957pspr0104_5}
}

@article{Anderson2012_3067,
  author = {Anderson, Cameron and Brion, Sebastien and Moore, Don A. and Kennedy, Jessica A.},
  publisher = {American Psychological Association},
  title = {{A status-enhancement account of overconfidence.}},
  journal = {Journal of Personality and Social Psychology},
  volume = {103},
  number = {4},
  pages = {718-735},
  year = {2012},
  doi = {10.1037/a0029395}
}

@article{Berger2014_3095,
  author = {Berger, Jonah},
  publisher = {John Wiley \& Sons, Ltd},
  title = {Word of Mouth and Interpersonal Communication: {A} Review and Directions for Future Research},
  journal = {Journal of Consumer Psychology},
  volume = {24},
  number = {4},
  pages = {586-607},
  year = {2014},
  doi = {10.1016/j.jcps.2014.05.002}
}

@article{Hsu2007_3079,
  author = {Hsu, Meng-Hsiang and Ju, Teresa L. and Yen, Chia-Hui and Chang, Chun-Ming},
  title = {Knowledge Sharing Behavior in Virtual Communities: {The} Relationship Between Trust, Self-Efficacy, and Outcome Expectations},
  journal = {International Journal of Human-Computer Studies},
  volume = {65},
  number = {2},
  pages = {153-169},
  year = {2007},
  doi = {10.1016/j.ijhcs.2006.09.003}
}

@article{Kim2020_3077,
  author = {Kim, Hannah and Lee, Jeongmin and Oh, Sung Eun},
  publisher = {Taylor \& Francis},
  title = {Individual Characteristics Influencing the Sharing of Knowledge on Social Networking Services: Online Identity, Self-Efficacy, and Knowledge Sharing Intentions},
  journal = {Behaviour \& Information Technology},
  volume = {39},
  number = {4},
  pages = {379-390},
  year = {2020},
  doi = {10.1080/0144929X.2019.1598494}
}

@article{Cheatham2017_3080,
  author = {Cheatham, Lauren B. and Tormala, Zakary L.},
  publisher = {SAGE Publications Inc},
  title = {The {Curvilinear} {Relationship} {Between} {Attitude} {Certainty} and {Attitudinal} {Advocacy}},
  journal = {Personality and Social Psychology Bulletin},
  volume = {43},
  number = {1},
  pages = {3-16},
  year = {2017},
  doi = {10.1177/0146167216673349}
}

@article{Bandiera2022_3096,
  author = {Bandiera, Oriana and Parekh, Nidhi and Petrongolo, Barbara and Rao, Michelle},
  publisher = {John Wiley \& Sons, Ltd},
  title = {Men Are from {Mars,} and {Women} {Too:} {A} {Bayesian} {Meta-analysis} of {Overconfidence} {Experiments}},
  journal = {Economica},
  volume = {89},
  number = {S1},
  pages = {S38-S70},
  year = {2022},
  doi = {10.1111/ecca.12407}
}

@article{Prims2017_3047,
  author = {Prims, Julia P. and Moore, Don A.},
  publisher = {Cambridge University Press},
  title = {Overconfidence over the Lifespan},
  journal = {Judgment and Decision Making},
  volume = {12},
  number = {1},
  pages = {29-41},
  year = {2017},
  doi = {10.1017/S1930297500005222}
}

@article{Ruisch2021_3104,
  author = {Ruisch, Benjamin C. and Stern, Chadly},
  publisher = {American Psychological Association},
  title = {{The confident conservative: Ideological differences in judgment and decision-making confidence.}},
  journal = {Journal of Experimental Psychology: General},
  volume = {150},
  number = {3},
  pages = {527-544},
  year = {2021},
  doi = {10.1037/xge0000898}
}

@article{Prooijen2019_3113,
  author = {van Prooijen, Jan-Willem and Krouwel, Andr\'{e} P. M.},
  publisher = {SAGE Publications Inc},
  title = {Psychological {Features} of {Extreme} {Political} {Ideologies}},
  journal = {Current Directions in Psychological Science},
  volume = {28},
  number = {2},
  pages = {159-163},
  year = {2019},
  doi = {10.1177/0963721418817755}
}

@article{Fernandes2026_3108,
  author = {Fernandes, Daniela and Villa, Steeven and Nicholls, Salla and Haavisto, Otso and Buschek, Daniel and Schmidt, Albrecht and Kosch, Thomas and Shen, Chenxinran and Welsch, Robin},
  title = {AI Makes You Smarter but None the Wiser: {The} Disconnect Between Performance and Metacognition},
  journal = {Computers in Human Behavior},
  volume = {175},
  pages = {108779},
  year = {2026},
  doi = {10.1016/j.chb.2025.108779}
}

@article{Sanchez2018_1208,
  author = {Sanchez, Carmen and Dunning, David},
  publisher = {American Psychological Association},
  title = {Overconfidence Among Beginners: {Is} a Little Learning a Dangerous Thing?},
  journal = {Journal of Personality and Social Psychology},
  volume = {114},
  pages = {10-28},
  year = {2018},
  address = {US},
  doi = {10.1037/pspa0000102}
}

@misc{Qualtrics2026_3020,
  author = {{Qualtrics}},
  title = {Fraud {Detection} - {Qualtrics}},
  year = {2026},
  url = {https://www.qualtrics.com/support/survey-platform/survey-module/survey-checker/fraud-detection/}
}

@article{Palmgreen2002_1970,
  author = {Palmgreen, Philip and Stephenson, Michael T. and Everett, Maureen W. and Baseheart, John R. and Francies, Regina},
  publisher = {Routledge},
  title = {Perceived {Message} {Sensation} {Value} {(PMSV)} and the {Dimensions} and {Validation} of a {PMSV} {Scale}},
  journal = {Health Communication},
  volume = {14},
  number = {4},
  pages = {403-428},
  year = {2002},
  doi = {10.1207/S15327027HC1404_1}
}

@article{Ho2017_3029,
  author = {Ho, Chin-Chang and MacDorman, Karl F.},
  title = {Measuring the {Uncanny} {Valley} {Effect}},
  journal = {International Journal of Social Robotics},
  volume = {9},
  number = {1},
  pages = {129-139},
  year = {2017},
  doi = {10.1007/s12369-016-0380-9}
}

@article{Wu2021_3027,
  author = {Wu, Linwan and Wen, Taylor Jing},
  publisher = {Taylor \& Francis},
  title = {Understanding {AI} {Advertising} {From} the {Consumer} {Perspective}},
  journal = {Journal of Advertising Research},
  volume = {61},
  number = {2},
  pages = {133-146},
  year = {2021},
  doi = {10.2501/JAR-2021-004}
}

@book{Robitzsch2025_3031,
  author = {Robitzsch, Alexander and Kiefer, Thomas and Wu, Margaret},
  title = {TAM: {Test} {Analysis} {Modules}},
  year = {2025},
  url = {https://CRAN.R-project.org/package=TAM},
  doi = {10.32614/CRAN.package.TAM}
}

@article{Stanislaw1999_3055,
  author = {Stanislaw, Harold and Todorov, Natasha},
  title = {Calculation of Signal Detection Theory Measures},
  journal = {Behavior Research Methods, Instruments, \& Computers},
  volume = {31},
  number = {1},
  pages = {137-149},
  year = {1999},
  doi = {10.3758/BF03207704}
}

@article{Hudgens2002_3037,
  author = {Hudgens, Michael and Satten, Glen},
  publisher = {Taylor \& Francis},
  title = {Midrank Unification of Rank Tests for Exact, Tied, and Censored Data},
  journal = {Journal of Nonparametric Statistics},
  volume = {14},
  number = {5},
  pages = {569-581},
  year = {2002},
  doi = {10.1080/10485250213905}
}

@article{Hyndman1996_3052,
  author = {Hyndman, Rob J. and Fan, Yanan},
  publisher = {Taylor \& Francis},
  title = {Sample {Quantiles} in {Statistical} {Packages}},
  journal = {The American Statistician},
  volume = {50},
  number = {4},
  pages = {361-365},
  year = {1996},
  doi = {10.1080/00031305.1996.10473566}
}

@article{Li2025_3039,
  author = {Li, Sophia and Hale, Randall and Moore, Don A.},
  publisher = {Cambridge University Press},
  title = {Is Overconfidence an Individual Difference?},
  journal = {Judgment and Decision Making},
  volume = {20},
  pages = {e25},
  year = {2025},
  doi = {10.1017/jdm.2025.11}
}

@article{Zhu2024_2336,
  author = {Zhu, Kayla},
  title = {Ranked: {The} {Most} {Popular} {Generative} {AI} {Tools} in 2024},
  journal = {Visual Capitalist},
  year = {2024},
  url = {https://www.visualcapitalist.com/ranked-the-most-popular-generative-ai-tools-in-2024/}
}

@article{Carolus2023_3033,
  author = {Carolus, Astrid and Koch, Martin J. and Straka, Samantha and Latoschik, Marc Erich and Wienrich, Carolin},
  title = {MAILS - {Meta} {AI} Literacy Scale: {Development} and Testing of an {AI} Literacy Questionnaire Based on Well-Founded Competency Models and Psychological Change- and Meta-Competencies},
  journal = {Computers in Human Behavior: Artificial Humans},
  volume = {1},
  number = {2},
  pages = {100014},
  year = {2023},
  doi = {10.1016/j.chbah.2023.100014}
}

@article{Qian2026_3034,
  author = {Qian, Sijia and Shen, Cuihua and Wang, Huiyi and Cho, Hichang},
  publisher = {SAGE Publications},
  title = {Building Resilience to Misinformation: {A} Cross-National Development of the {Digital} {Media} and {Information} {Literacy} {Scale} {(DMILS)}},
  journal = {New Media \& Society},
  year = {2026},
  url = {https://doi.org/10.1177/14614448261448549},
  doi = {10.1177/14614448261448549}
}

@article{Hoyle2002_1680,
  author = {Hoyle, Rick H and Stephenson, Michael T. and Palmgreen, Philip and Lorch, Elizabeth Pugzles and Donohew, R.Lewis},
  title = {Reliability and Validity of a Brief Measure of Sensation Seeking},
  journal = {Personality and Individual Differences},
  volume = {32},
  number = {3},
  pages = {401-414},
  year = {2002},
  doi = {10.1016/S0191-8869(01)00032-0}
}

@article{Barasch2014_3022,
  author = {Barasch, Alixandra and Berger, Jonah},
  publisher = {SAGE Publications Inc},
  title = {Broadcasting and {Narrowcasting:} {How} {Audience} {Size} {Affects} {What} {People} {Share}},
  journal = {Journal of Marketing Research},
  volume = {51},
  number = {3},
  pages = {286-299},
  year = {2014},
  doi = {10.1509/jmr.13.0238}
}

@article{Kuznetsova2017_99,
  author = {Kuznetsova, Alexandra and Brockhoff, Per B. and Christensen, Rune H. B.},
  publisher = {Foundation for Open Access Statistic},
  title = {lmerTest {Package:} {Tests} in {Linear} {Mixed} {Effects} {Models}},
  journal = {Journal of Statistical Software},
  volume = {82},
  number = {13},
  pages = {1-26},
  year = {2017},
  doi = {10.18637/JSS.V082.I13}
}

@article{Kruger1999_27,
  author = {Kruger, Justin and Dunning, David},
  publisher = {American Psychological Association (APA)},
  title = {Unskilled and Unaware of It: {How} Difficulties in Recognizing One's Own Incompetence Lead to Inflated Self-Assessments.},
  journal = {Journal of Personality and Social Psychology},
  volume = {77},
  number = {6},
  pages = {1121-1134},
  year = {1999},
  doi = {10.1037//0022-3514.77.6.1121}
}

@article{Thompson2011_3115,
  author = {Thompson, Valerie A. and Prowse Turner, Jamie A. and Pennycook, Gordon},
  title = {Intuition, Reason, and Metacognition},
  journal = {Cognitive Psychology},
  volume = {63},
  number = {3},
  pages = {107-140},
  year = {2011},
  doi = {10.1016/j.cogpsych.2011.06.001}
}

@article{Lim2026_3122,
  author = {Lim, Han Sol and Lee, Bong Gyou and Sung, Yoon Hi and Jung, Chang Won},
  title = {Human-Made Vs. {AI-generated:} How Provenance Labels Drive Strategic Curation via Perceived Effort},
  journal = {Frontiers in Psychology},
  volume = {17},
  year = {2026},
  doi = {10.3389/fpsyg.2026.1840483}
}

@article{Raj2026_3118,
  author = {Raj, Manav and Berg, Justin M. and Seamans, Rob},
  publisher = {American Psychological Association},
  title = {{The artificial intelligence disclosure penalty: Humans persistently devalue AI-generated creative writing.}},
  journal = {Journal of Experimental Psychology: General},
  volume = {155},
  number = {4},
  pages = {896-915},
  year = {2026},
  doi = {10.1037/xge0001889}
}

@article{Chung2026_3125,
  author = {Chung, Myojung},
  publisher = {SAGE Publications},
  title = {Overconfident Youth, Underconfident Elders: {Age} Differences in Algorithmic Knowledge Confidence and Information Perceptions},
  journal = {New Media \& Society},
  pages = {14614448261459666},
  year = {2026},
  doi = {10.1177/14614448261459666}
}

@article{Grzenkowicz2025_3026,
  author = {Grzenkowicz, Maciej and Wildfeuer, Janina},
  title = {Addressing {TikTok's} Multimodal Complexity: A Multi-Level Annotation Scheme for the Audio-Visual Design of Short Video Content},
  journal = {Digital Scholarship in the Humanities},
  volume = {40},
  number = {4},
  pages = {1143-1166},
  year = {2025},
  doi = {10.1093/llc/fqaf047}
}

@article{Mair2007_3030,
  author = {Mair, Patrick and Hatzinger, Reinhold},
  title = {Extended {Rasch} {Modeling:} {The} {eRm} {Package} for the {Application} of {IRT} {Models} in {R}},
  journal = {Journal of Statistical Software},
  volume = {20},
  number = {9},
  pages = {1 - 20},
  year = {2007},
  doi = {10.18637/jss.v020.i09}
}
\clearpage
\section*{Supplemental Information}
\addcontentsline{toc}{section}{Supplemental Information}
\setcounter{figure}{0}
\renewcommand{\thefigure}{S\arabic{figure}}
\setcounter{table}{0}
\renewcommand{\thetable}{S\arabic{table}}
\renewcommand{\theHtable}{S\arabic{table}}
\renewcommand{\theHfigure}{S\arabic{figure}}
\section*{Supplemental Methods: Stimulus Development and Pilot Testing}

\subsection*{S1. Video Creation}

Before creating videos, we selected topics that represent common social media content within each context to ensure external validity, including nine health topics (e.g., healthy eating, physical activity), eight environmental topics (e.g., wildfire, endangered animals), and nine entertainment topics (e.g., sports, dancing).

We used three popular generative AI tools to create GenAI stimuli: OpenAI's Sora 2, Google's Veo 3, and Midjourney. We selected these tools based on their capacity to generate realistic videos and availability during the stimuli development stage. For each topic, we started by identifying a specific scene, such as preparing avocado toast to represent healthy eating. We then developed topic-specific prompts using a consistent structure, specifying the subject and action, setting, and production characteristics (e.g., aspect ratio, visual quality, camera framing, angle, and lighting). These characteristics varied across topics to mimic typical short-video formats (e.g., handheld recordings, documentary footage, and lifestyle clips) rather than consistently polished footage \citep{Grzenkowicz2025_3026}. For example, an exercise video used steady framing and natural lighting, whereas a hurricane video featured lower-quality handheld footage taken through a car windshield. This approach added visual variety while keeping each scene relevant to its topic.

We applied the same topic-specific prompts across all three tools to produce candidate videos. Unlike the other two prompt-to-video tools, Midjourney first generated a starting image from the prompt, then animated it into a video clip. During this process, we refined the prompts as needed when the output consistently contained implausible motion, discontinuities, or other critical visual inconsistencies. The research team reviewed the remaining videos for visual realism and topic relevance, excluding videos with an artificial overall appearance, such as unnatural textures or persistent visual distortions. However, screening did not require videos to be free from detectable errors. We kept some videos with localized physical inconsistencies to maintain variation in detection difficulty.

Next, we selected human-made counterparts for each AI-generated video from publicly available social media posts, advertisements, documentaries, and news footage on TikTok, YouTube, and news websites. We prioritized comparability with the AI-generated videos in topic, depicted subjects and actions, and visual presentation. We reviewed source attribution and available provenance information to determine that the footage was human-made. We trimmed human-made clips to lengths comparable to those of AI-generated videos, which typically lasted 5 to 10 seconds. All videos were presented without audio to standardize presentation across stimuli. This was because audio availability and quality varied: Midjourney-generated stimuli did not include sound, but other AI-generated videos often showed obvious audiovisual mismatches. Human-made videos included different combinations of background music, speech, and ambient sound. Removing audio therefore allowed us to focus on visual discernment without additional auditory cues.

\subsection*{S2. Pilot Testing and Stimuli Selection}

We pilot-tested 58 candidate videos (32 AI-generated and 26 human-made videos) on \emph{n} = 100 participants recruited on CloudResearch Connect. This pilot aimed to assess (a) variation in detection difficulty and identify unsuitable stimuli and (b) the comparability of AI-generated and human-made videos in subjective sensory experiences. Every participant evaluated 12 random videos, four from each context. On average, each video received 20.69 evaluations (\emph{SD} = 6.16). For each video, participants classified its source and evaluated perceived message sensation value (PMSV; \citep{Palmgreen2002_1970}) and perceived eeriness \citep{Ho2017_3029,Wu2021_3027} on a seven-point scale. See Tables~\ref{tab:s2}--\ref{tab:s3} for measurement details. We aggregated responses by video to calculate classification accuracy and perception ratings.

First, we used Rasch modeling to estimate video difficulty using the \emph{TAM} package in R \citep{Robitzsch2025_3031} and evaluate model fit (i.e., outfit mean-square statistics) using the \emph{eRm} package \citep{Mair2007_3030}. Estimated difficulty reflects item difficulty while accounting for individual differences; scores within \ensuremath{\pm}2 logits are considered acceptable. Outfit statistics identify unexpected classifications, such as a high-ability participant incorrectly classifying an otherwise easy video; statistics within 0.5 and 1.5 are considered acceptable. These measures informed video selection with suitable difficulty and response patterns. If no suitable AI-generated or human-made version remained for a topic after this process, we excluded the entire topic. This process produced 42 videos, organized into 21 topic pairs of AI-generated and human-made videos, with seven topic pairs in each context.

The final stimulus set had a mean duration of 6.97 seconds (\emph{SD} = 1.17, \emph{Mdn} = 7.06, \emph{Min} = 5.21, \emph{Max} = 9.31). Classification accuracy varied across videos, with an average accuracy of 59.94\% (\emph{SD} = 15.99\%, \emph{Min} = 25\%, \emph{Max} = 88\%). Average PMSV (\emph{M} = 2.99, \emph{SD} = 0.59) and perceived eeriness (\emph{M} = 3.17, \emph{SD} = 0.49) were both below the midpoint of their seven-point scale. We also ran linear mixed-effects models and found no significant source difference in PMSV (\emph{b} = 0.23, \emph{SE} = 0.54, \emph{p} = .676) or perceived eeriness (\emph{b} = -0.17, \emph{SE} = 0.43, \emph{p} = .689). See Table~\ref{tab:s4} for summary statistics. Overall, this pilot showed variation in detection difficulty but no statistically significant differences in PMSV or perceived eeriness across sources.

\clearpage
\begin{landscape}
\begin{figure}[p]
\centering
\includegraphics[width=\linewidth,height=.80\textheight,keepaspectratio]{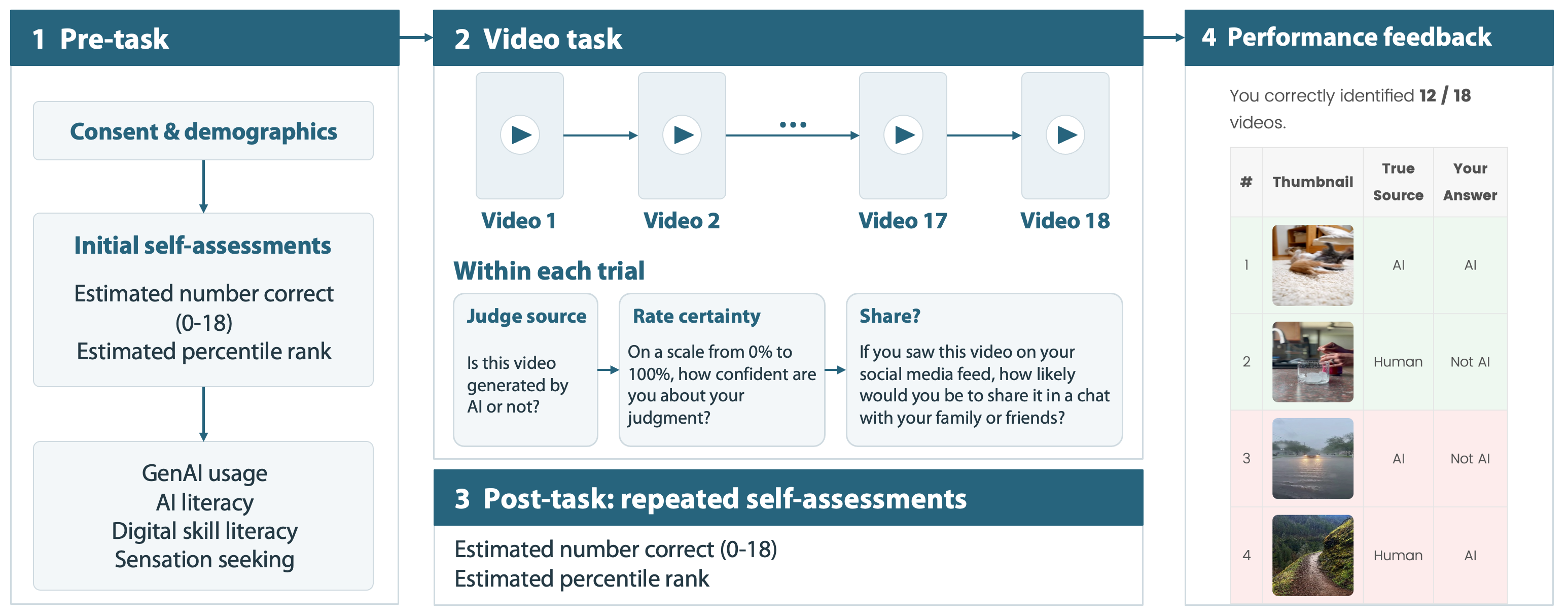}
\caption{Illustration of study procedures.}\label{fig:s1}
\end{figure}
\clearpage
\end{landscape}

\clearpage
\begin{figure}[p]
\centering
\includegraphics[width=\linewidth,height=.80\textheight,keepaspectratio]{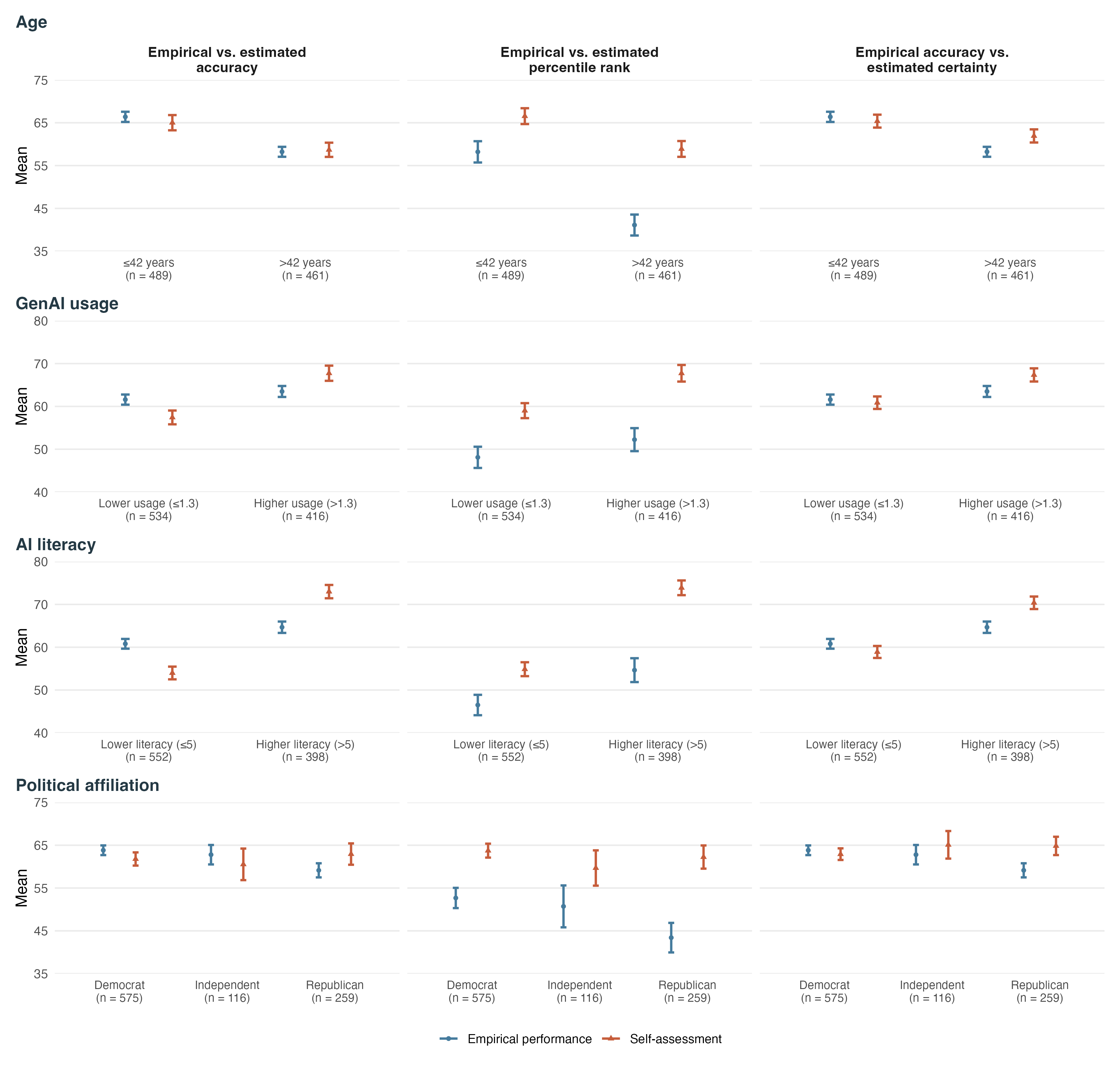}
\caption{Average actual performance and self-assessments by age, GenAI usage, AI literacy, and political affiliation. Points show unadjusted group means, with 95\% confidence intervals. Age, GenAI usage, and AI literacy were split at their sample medians.}\label{fig:s2}
\end{figure}
\clearpage

\clearpage
\begingroup
\small
\setlength{\tabcolsep}{4pt}
\renewcommand{\arraystretch}{1.12}
\captionof{table}{Summary statistics of participant demographics (\emph{N} = 973).}\label{tab:s1}
{ % do not increment counter
\begin{longtable*}[]{@{}
  >{\raggedright\arraybackslash}p{(\linewidth - 6\tabcolsep) * \real{0.1818}}
  >{\raggedright\arraybackslash}p{(\linewidth - 6\tabcolsep) * \real{0.5604}}
  >{\centering\arraybackslash}p{(\linewidth - 6\tabcolsep) * \real{0.1289}}
  >{\centering\arraybackslash}p{(\linewidth - 6\tabcolsep) * \real{0.1289}}@{}}
\toprule\noalign{}
\begin{minipage}[b]{\linewidth}\raggedright
\textbf{Characteristic}
\end{minipage} & \begin{minipage}[b]{\linewidth}\raggedright
\textbf{Category}
\end{minipage} & \begin{minipage}[b]{\linewidth}\centering
\textbf{n}
\end{minipage} & \begin{minipage}[b]{\linewidth}\centering
\textbf{\%}
\end{minipage} \\
\midrule\noalign{}
\endhead
\bottomrule\noalign{}
\endlastfoot
Age & M (SD) & \multicolumn{2}{>{\centering\arraybackslash}p{(\linewidth - 6\tabcolsep) * \real{0.2579} + 2\tabcolsep}@{}}{%
43.39 (15.31)} \\
Gender & Female & 478 & 49.5 \\
& Male & 487 & 50.4 \\
& Non-binary & 1 & 0.1 \\
Race & Asian & 84 & 8.6 \\
& Black & 125 & 12.8 \\
& American Indian/Native American or Alaska Native & 14 & 1.4 \\
& Native Hawaiian or Other Pacific Islander & 1 & 0.1 \\
& White & 712 & 73.2 \\
& Hispanic & 121 & 12.4 \\
Education & Some high school or less & 6 & 0.6 \\
& High School/GED & 117 & 12.1 \\
& Some college no degree & 205 & 21.2 \\
& Associates/technical degree & 116 & 12.0 \\
& Bachelor's degree & 359 & 37.1 \\
& Graduate/professional degree & 164 & 17.0 \\
Income & \textless\$25,000 & 111 & 11.6 \\
& \$25,000--\$49,999 & 222 & 23.2 \\
& \$50,000--\$74,999 & 196 & 20.5 \\
& \$75,000--\$99,999 & 160 & 16.8 \\
& \$100,000--\$149,999 & 172 & 18.0 \\
& \ensuremath{\geq}\$150,000 & 94 & 9.8 \\
Party affiliation & Democrat & 586 & 60.2 \\
& Independent & 127 & 13.1 \\
& Republican & 260 & 26.7 \\
\end{longtable*}
}
\endgroup

\clearpage
\begingroup
\small
\setlength{\tabcolsep}{4pt}
\renewcommand{\arraystretch}{1.12}
\captionof{table}{Measurement of perceived message sensation value using five items selected from the original 17-item scale.}\label{tab:s2}
{ % do not increment counter
\begin{longtable*}[]{@{}
  >{\raggedright\arraybackslash}p{(\linewidth - 2\tabcolsep) * \real{0.3929}}
  >{\raggedright\arraybackslash}p{(\linewidth - 2\tabcolsep) * \real{0.6071}}@{}}
\toprule\noalign{}
\begin{minipage}[b]{\linewidth}\raggedright
\textbf{Original scale dimension}
\end{minipage} & \begin{minipage}[b]{\linewidth}\raggedright
\textbf{Items (1-7)}
\end{minipage} \\
\midrule\noalign{}
\endhead
\bottomrule\noalign{}
\endlastfoot
Emotional arousal & Emotional--unemotional

Involving--uninvolving \\
Dramatic impact & Dramatic--undramatic

Intense--not intense \\
Novelty & Novel--ordinary \\
\end{longtable*}
}
\endgroup

\par\bigskip
\begingroup
\small
\setlength{\tabcolsep}{4pt}
\renewcommand{\arraystretch}{1.12}
\captionof{table}{Measurement of perceived eeriness.}\label{tab:s3}
{ % do not increment counter
\begin{longtable*}[]{@{}
  >{\raggedright\arraybackslash}p{(\linewidth - 2\tabcolsep) * \real{0.7077}}
  >{\centering\arraybackslash}p{(\linewidth - 2\tabcolsep) * \real{0.2923}}@{}}
\toprule\noalign{}
\begin{minipage}[b]{\linewidth}\raggedright
\textbf{Items}
\end{minipage} & \begin{minipage}[b]{\linewidth}\centering
\textbf{Response options}
\end{minipage} \\
\midrule\noalign{}
\endhead
\bottomrule\noalign{}
\endlastfoot
Creepy--Not creepy & \multirow{9}{=}{\centering\arraybackslash Seven-point semantic differential} \\
Artificial--Natural \\
Weird--Normal \\
Bizarre--Familiar \\
Attractive--Unattractive \\
Synthetic--Real \\
Freaky--Usual \\
Spine-tingling--Dull \\
Low production quality--High production quality \\
\end{longtable*}
}
\endgroup

\clearpage
\begin{landscape}
\begingroup
\small
\setlength{\tabcolsep}{4pt}
\renewcommand{\arraystretch}{1.12}
\captionof{table}{Descriptive statistics of the final stimulus set by video source and context (\emph{n} = 42 videos).}\label{tab:s4}
{ % do not increment counter
\begin{longtable*}[]{@{}
  >{\raggedright\arraybackslash}p{(\linewidth - 12\tabcolsep) * \real{0.1204}}
  >{\centering\arraybackslash}p{(\linewidth - 12\tabcolsep) * \real{0.1203}}
  >{\centering\arraybackslash}p{(\linewidth - 12\tabcolsep) * \real{0.0547}}
  >{\centering\arraybackslash}p{(\linewidth - 12\tabcolsep) * \real{0.1203}}
  >{\centering\arraybackslash}p{(\linewidth - 12\tabcolsep) * \real{0.1860}}
  >{\centering\arraybackslash}p{(\linewidth - 12\tabcolsep) * \real{0.2078}}
  >{\centering\arraybackslash}p{(\linewidth - 12\tabcolsep) * \real{0.1906}}@{}}
\toprule\noalign{}
\begin{minipage}[b]{\linewidth}\raggedright
\textbf{Context}
\end{minipage} & \begin{minipage}[b]{\linewidth}\centering
\textbf{Source}
\end{minipage} & \begin{minipage}[b]{\linewidth}\centering
\textbf{n}
\end{minipage} & \begin{minipage}[b]{\linewidth}\centering
\textbf{Duration (s)\\
M (SD)}\strut
\end{minipage} & \begin{minipage}[b]{\linewidth}\centering
\textbf{Accuracy (\%)\\
M (SD)}\strut
\end{minipage} & \begin{minipage}[b]{\linewidth}\centering
\textbf{Perceived message sensation value\\
M (SD)}\strut
\end{minipage} & \begin{minipage}[b]{\linewidth}\centering
\textbf{Perceived eeriness\\
M (SD)}\strut
\end{minipage} \\
\midrule\noalign{}
\endhead
\bottomrule\noalign{}
\endlastfoot
- & Human-made & 21 & 6.84 (1.03) & 61.01 (17.32) & 2.96 (0.65) & 3.07 (0.48) \\
- & AI-generated & 21 & 7.10 (1.31) & 58.87 (14.89) & 3.01 (0.55) & 3.26 (0.50) \\
Entertainment & - & 14 & 6.95 (1.04) & 59.27 (15.49) & 3.01 (0.52) & 2.94 (0.36) \\
Health & - & 14 & 6.67 (1.33) & 60.86 (15.33) & 2.52 (0.33) & 3.16 (0.48) \\
Environment & - & 14 & 7.29 (1.13) & 59.68 (18.17) & 3.43 (0.55) & 3.39 (0.55) \\
\multirow{2}{=}{Entertainment} & Human-made & 7 & 6.68 (0.55) & 57.75 (18.64) & 3.04 (0.57) & 2.87 (0.35) \\
& AI-generated & 7 & 7.22 (1.37) & 60.79 (12.92) & 2.98 (0.51) & 3.02 (0.37) \\
\multirow{2}{=}{Health} & Human-made & 7 & 6.79 (1.31) & 61.57 (17.07) & 2.46 (0.34) & 2.98 (0.53) \\
& AI-generated & 7 & 6.55 (1.44) & 60.15 (14.71) & 2.58 (0.34) & 3.34 (0.36) \\
\multirow{2}{=}{Environment} & Human-made & 7 & 7.04 (1.21) & 63.70 (18.44) & 3.38 (0.68) & 3.36 (0.45) \\
& AI-generated & 7 & 7.53 (1.07) & 55.66 (18.38) & 3.47 (0.42) & 3.42 (0.66) \\
\multicolumn{2}{@{}>{\centering\arraybackslash}p{(\linewidth - 12\tabcolsep) * \real{0.2407} + 2\tabcolsep}}{%
Combined} & 42 & 6.97 (1.17) & 59.94 (15.99) & 2.99 (0.59) & 3.17 (0.49) \\
\end{longtable*}
}
\endgroup
\end{landscape}

\clearpage
\begingroup
\small
\setlength{\tabcolsep}{4pt}
\renewcommand{\arraystretch}{1.12}
\captionof{table}{Pearson correlations among overestimation, overplacement, and overprecision at the individual level (\emph{N} = 973).}\label{tab:s5}
{ % do not increment counter
\begin{longtable*}[]{@{}
  >{\raggedright\arraybackslash}p{(\linewidth - 6\tabcolsep) * \real{0.3076}}
  >{\centering\arraybackslash}p{(\linewidth - 6\tabcolsep) * \real{0.2308}}
  >{\centering\arraybackslash}p{(\linewidth - 6\tabcolsep) * \real{0.2308}}
  >{\centering\arraybackslash}p{(\linewidth - 6\tabcolsep) * \real{0.2308}}@{}}
\toprule\noalign{}
\begin{minipage}[b]{\linewidth}\raggedright
\textbf{Measure}
\end{minipage} & \begin{minipage}[b]{\linewidth}\centering
\textbf{Overestimation}
\end{minipage} & \begin{minipage}[b]{\linewidth}\centering
\textbf{Overplacement}
\end{minipage} & \begin{minipage}[b]{\linewidth}\centering
\textbf{Overprecision}
\end{minipage} \\
\midrule\noalign{}
\endhead
\bottomrule\noalign{}
\endlastfoot
Overestimation & --- & & \\
Overplacement & 0.78 & --- & \\
Overprecision & 0.65 & 0.65 & --- \\
\end{longtable*}
}
\endgroup

\par\bigskip
\begingroup
\small
\setlength{\tabcolsep}{4pt}
\renewcommand{\arraystretch}{1.12}
\captionof{table}{Generative AI tools included in the measure of GenAI usage frequency.}\label{tab:s6}
{ % do not increment counter
\begin{longtable*}[]{@{}
  >{\raggedright\arraybackslash}p{(\linewidth - 2\tabcolsep) * \real{0.7077}}
  >{\raggedright\arraybackslash}p{(\linewidth - 2\tabcolsep) * \real{0.2923}}@{}}
\toprule\noalign{}
\begin{minipage}[b]{\linewidth}\raggedright
\textbf{Generative AI tool}
\end{minipage} & \begin{minipage}[b]{\linewidth}\raggedright
\textbf{Response options}
\end{minipage} \\
\midrule\noalign{}
\endhead
\bottomrule\noalign{}
\endlastfoot
ChatGPT & \multirow{10}{=}{\begin{minipage}[t]{\linewidth}\raggedright
1 = Never\\
2 = Sometimes\\
3 = About half the time\\
4 = Most of the time\\
5 = Always\strut
\end{minipage}} \\
Nano Banana \\
Midjourney \\
Reve \\
Ideogram \\
Google Veo \\
Runway \\
Sora \\
Descript \\
Wondershare Filmora \\
\end{longtable*}
}
\endgroup

\clearpage
\begingroup
\small
\setlength{\tabcolsep}{4pt}
\renewcommand{\arraystretch}{1.12}
\captionof{table}{Measurement of AI literacy.}\label{tab:s7}
{ % do not increment counter
\begin{longtable*}[]{@{}
  >{\raggedright\arraybackslash}p{(\linewidth - 2\tabcolsep) * \real{0.7230}}
  >{\centering\arraybackslash}p{(\linewidth - 2\tabcolsep) * \real{0.2770}}@{}}
\toprule\noalign{}
\begin{minipage}[b]{\linewidth}\raggedright
\textbf{Items}
\end{minipage} & \begin{minipage}[b]{\linewidth}\centering
\textbf{Response options}
\end{minipage} \\
\midrule\noalign{}
\endhead
\bottomrule\noalign{}
\endlastfoot
I can tell if I am dealing with an application based on artificial intelligence. & \multirow{3}{=}{\centering\arraybackslash \begin{minipage}[t]{\linewidth}\centering
1 = Strongly disagree\\
7 = Strongly agree\strut
\end{minipage}} \\
I can distinguish devices that use AI from devices that do not. \\
I can distinguish if I interact with an AI or a ''real human''. \\
\end{longtable*}
}
\endgroup

\par\bigskip
\begingroup
\small
\setlength{\tabcolsep}{4pt}
\renewcommand{\arraystretch}{1.12}
\captionof{table}{Measurement of digital skill literacy.}\label{tab:s8}
{ % do not increment counter
\begin{longtable*}[]{@{}
  >{\raggedright\arraybackslash}p{(\linewidth - 2\tabcolsep) * \real{0.7230}}
  >{\centering\arraybackslash}p{(\linewidth - 2\tabcolsep) * \real{0.2770}}@{}}
\toprule\noalign{}
\begin{minipage}[b]{\linewidth}\raggedright
\textbf{Items}
\end{minipage} & \begin{minipage}[b]{\linewidth}\centering
\textbf{Response options}
\end{minipage} \\
\midrule\noalign{}
\endhead
\bottomrule\noalign{}
\endlastfoot
I know how to solve my own technical problems on digital devices and platforms. & \multirow{4}{=}{\centering\arraybackslash \begin{minipage}[t]{\linewidth}\centering
1 = Strongly disagree\\
7 = Strongly agree\strut
\end{minipage}} \\
I can learn new technologies easily. \\
I have the technical skills I need to use information and communication technologies (ICT) for learning and to create artifacts (e.g., presentations, digital stories, wikis, blogs). \\
Using any technological device comes easy to me. \\
\end{longtable*}
}
\endgroup

\par\bigskip
\begingroup
\small
\setlength{\tabcolsep}{4pt}
\renewcommand{\arraystretch}{1.12}
\captionof{table}{Measurement of sensation seeking.}\label{tab:s9}
{ % do not increment counter
\begin{longtable*}[]{@{}
  >{\centering\arraybackslash}p{(\linewidth - 2\tabcolsep) * \real{0.0601}}
  >{\raggedright\arraybackslash}p{(\linewidth - 2\tabcolsep) * \real{0.9399}}@{}}
\toprule\noalign{}
\begin{minipage}[b]{\linewidth}\centering
\textbf{\#}
\end{minipage} & \begin{minipage}[b]{\linewidth}\centering
\textbf{Question item}
\end{minipage} \\
\midrule\noalign{}
\endhead
\bottomrule\noalign{}
\endlastfoot
1 & I would like to explore strange places. \\
2 & I would like to take off on a trip with no pre-planned routes or timetables. \\
3 & I get restless when I spend too much time at home. \\
4 & I prefer friends who are excitingly unpredictable. \\
5 & I like to do frightening things. \\
6 & I would like to try bungee jumping. \\
7 & I like wild parties. \\
8 & I would love to have new and exciting experiences, even if they are illegal. \\
\end{longtable*}
}
\endgroup

\clearpage
\begin{landscape}
\begingroup
\small
\setlength{\tabcolsep}{4pt}
\renewcommand{\arraystretch}{1.12}
\captionof{table}{Descriptive statistics of trial-level responses by video source and context in the main study.}\label{tab:s10}
{ % do not increment counter
\begin{longtable*}[]{@{}
  >{\raggedright\arraybackslash}p{(\linewidth - 10\tabcolsep) * \real{0.1131}}
  >{\centering\arraybackslash}p{(\linewidth - 10\tabcolsep) * \real{0.1357}}
  >{\centering\arraybackslash}p{(\linewidth - 10\tabcolsep) * \real{0.2262}}
  >{\centering\arraybackslash}p{(\linewidth - 10\tabcolsep) * \real{0.1244}}
  >{\centering\arraybackslash}p{(\linewidth - 10\tabcolsep) * \real{0.2036}}
  >{\centering\arraybackslash}p{(\linewidth - 10\tabcolsep) * \real{0.1970}}@{}}
\toprule\noalign{}
\begin{minipage}[b]{\linewidth}\raggedright
\textbf{Context}
\end{minipage} & \begin{minipage}[b]{\linewidth}\centering
\textbf{Source}
\end{minipage} & \begin{minipage}[b]{\linewidth}\centering
\textbf{Perceived source (\% classified as human-made)\\
M (SD)}\strut
\end{minipage} & \begin{minipage}[b]{\linewidth}\centering
\textbf{Accuracy (\%)\\
M (SD)}\strut
\end{minipage} & \begin{minipage}[b]{\linewidth}\centering
\textbf{Estimated certainty (\%)\\
M (SD)}\strut
\end{minipage} & \begin{minipage}[b]{\linewidth}\centering
\textbf{Sharing intention\\
M (SD)}\strut
\end{minipage} \\
\midrule\noalign{}
\endhead
\bottomrule\noalign{}
\endlastfoot
- & Human-made & 63.45 (19.46) & 63.45 (19.46) & 62.04 (18.05) & 2.48 (1.39) \\
- & AI-generated & 38.49 (20.91) & 61.51 (20.91) & 65.37 (17.37) & 2.39 (1.40) \\
Entertainment & - & 52.21 (21.54) & 65.86 (20.94) & 64.10 (18.09) & 2.63 (1.48) \\
Health & - & 47.17 (21.63) & 63.81 (20.37) & 64.96 (18.12) & 2.09 (1.33) \\
Environment & - & 53.53 (23.05) & 57.76 (20.34) & 62.05 (17.86) & 2.59 (1.49) \\
\multirow{2}{=}{Entertainment} & Human-made & 68.07 (28.55) & 68.07 (28.55) & 62.03 (20.25) & 2.72 (1.59) \\
& AI-generated & 36.35 (31.46) & 63.65 (31.46) & 66.17 (19.10) & 2.54 (1.57) \\
\multirow{2}{=}{Health} & Human-made & 60.98 (29.55) & 60.98 (29.55) & 63.00 (19.92) & 2.17 (1.42) \\
& AI-generated & 33.37 (29.88) & 66.63 (29.88) & 66.93 (19.12) & 2.00 (1.36) \\
\multirow{2}{=}{Environment} & Human-made & 61.29 (30.87) & 61.29 (30.87) & 61.08 (19.07) & 2.55 (1.53) \\
& AI-generated & 45.77 (30.61) & 54.23 (30.61) & 63.02 (18.93) & 2.63 (1.60) \\
\multicolumn{2}{@{}>{\centering\arraybackslash}p{(\linewidth - 10\tabcolsep) * \real{0.2488} + 2\tabcolsep}}{%
Combined} & 50.97 (14.85) & 62.48 (13.69) & 63.71 (17.07) & 2.44 (1.36) \\
\end{longtable*}
}
\endgroup
\end{landscape}

\clearpage
\begingroup
\small
\setlength{\tabcolsep}{4pt}
\renewcommand{\arraystretch}{1.12}
\captionof{table}{Estimated confidence and actual performance across the accuracy groups shown in Figure~\ref{fig:2} (\emph{N} = 973). Participants were grouped into four accuracy quartiles (0-9, 10-11, 12-13, and 14-18 correct classifications out of 18); ties are preserved.}\label{tab:s11}
{ % do not increment counter
\begin{longtable*}[]{@{}
  >{\raggedright\arraybackslash}p{(\linewidth - 6\tabcolsep) * \real{0.3076}}
  >{\raggedright\arraybackslash}p{(\linewidth - 6\tabcolsep) * \real{0.0923}}
  >{\raggedright\arraybackslash}p{(\linewidth - 6\tabcolsep) * \real{0.3000}}
  >{\raggedright\arraybackslash}p{(\linewidth - 6\tabcolsep) * \real{0.3000}}@{}}
\toprule\noalign{}
\begin{minipage}[b]{\linewidth}\raggedright
\textbf{Accuracy group}
\end{minipage} & \begin{minipage}[b]{\linewidth}\centering
\textbf{n}
\end{minipage} & \begin{minipage}[b]{\linewidth}\centering
\textbf{Estimated M (SD)}
\end{minipage} & \begin{minipage}[b]{\linewidth}\centering
\textbf{Actual M (SD)}
\end{minipage} \\
\midrule\noalign{}
\endhead
\bottomrule\noalign{}
\endlastfoot
\multicolumn{4}{@{}>{\raggedright\arraybackslash}p{(\linewidth - 6\tabcolsep) * \real{1.0000} + 6\tabcolsep}@{}}{%
\textbf{Overestimation: estimated and actual accuracy (\%)}} \\
Lowest & 247 & 58.84 (19.41) & 44.87 (5.92) \\
Lower-middle & 275 & 60.85 (19.15) & 58.53 (2.78) \\
Upper-middle & 259 & 64.20 (18.91) & 69.07 (2.76) \\
Highest & 192 & 64.38 (19.31) & 81.89 (5.19) \\
\multicolumn{4}{@{}>{\raggedright\arraybackslash}p{(\linewidth - 6\tabcolsep) * \real{1.0000} + 6\tabcolsep}@{}}{%
\textbf{Overplacement: estimated and actual percentile rank}} \\
Lowest & 247 & 59.44 (20.36) & 12.69 (6.87) \\
Lower-middle & 275 & 60.16 (21.89) & 39.52 (7.06) \\
Upper-middle & 259 & 66.04 (19.43) & 66.96 (6.61) \\
Highest & 192 & 66.66 (20.42) & 90.13 (5.21) \\
\multicolumn{4}{@{}>{\raggedright\arraybackslash}p{(\linewidth - 6\tabcolsep) * \real{1.0000} + 6\tabcolsep}@{}}{%
\textbf{Overprecision: mean judgment confidence and actual accuracy (\%)}} \\
Lowest & 247 & 62.07 (17.27) & 44.87 (5.92) \\
Lower-middle & 275 & 61.57 (18.06) & 58.53 (2.78) \\
Upper-middle & 259 & 65.12 (16.05) & 69.07 (2.76) \\
Highest & 192 & 66.96 (16.12) & 81.89 (5.19) \\
\end{longtable*}
}
\endgroup

\clearpage
\begingroup
\small
\setlength{\tabcolsep}{4pt}
\renewcommand{\arraystretch}{1.12}
\captionof{table}{Individual-level linear regression models predicting response tendency from overestimation, overplacement, and overprecision (\emph{n} = 950). Numeric predictors are standardized.}\label{tab:s12}
{ % do not increment counter
\begin{longtable*}[]{@{}
  >{\raggedright\arraybackslash}p{(\linewidth - 8\tabcolsep) * \real{0.4001}}
  >{\centering\arraybackslash}p{(\linewidth - 8\tabcolsep) * \real{0.2031}}
  >{\centering\arraybackslash}p{(\linewidth - 8\tabcolsep) * \real{0.0969}}
  >{\centering\arraybackslash}p{(\linewidth - 8\tabcolsep) * \real{0.2031}}
  >{\centering\arraybackslash}p{(\linewidth - 8\tabcolsep) * \real{0.0969}}@{}}
\toprule\noalign{}
\begin{minipage}[b]{\linewidth}\raggedright
\textbf{Predictor}
\end{minipage} & \multicolumn{2}{>{\centering\arraybackslash}p{(\linewidth - 8\tabcolsep) * \real{0.3000} + 2\tabcolsep}}{%
\begin{minipage}[b]{\linewidth}\centering
\textbf{Model 1}
\end{minipage}} & \multicolumn{2}{>{\centering\arraybackslash}p{(\linewidth - 8\tabcolsep) * \real{0.3000} + 2\tabcolsep}@{}}{%
\begin{minipage}[b]{\linewidth}\centering
\textbf{Model 2}
\end{minipage}} \\
\begin{minipage}[b]{\linewidth}\raggedright
\end{minipage} & \begin{minipage}[b]{\linewidth}\centering
\textbf{b (SE)}
\end{minipage} & \begin{minipage}[b]{\linewidth}\centering
\textbf{p}
\end{minipage} & \begin{minipage}[b]{\linewidth}\centering
\textbf{b (SE)}
\end{minipage} & \begin{minipage}[b]{\linewidth}\centering
\textbf{p}
\end{minipage} \\
\midrule\noalign{}
\endhead
\bottomrule\noalign{}
\endlastfoot
Intercept & 0.03 (0.01) & .055 & 0.05 (0.03) & .099 \\
Overestimation & -0.04 (0.02) & .100 & -0.02 (0.02) & .529 \\
Overplacement & 0.02 (0.02) & .394 & 0.01 (0.02) & .644 \\
Overprecision & 0.04 (0.02)* & .024 & 0.04 (0.02)* & .033 \\
Age & & & 0.010 (0.016) & .541 \\
Gender: female & & & -0.05 (0.03)* & .045 \\
Race: White & & & 0.005 (0.033) & .873 \\
Education & & & -0.007 (0.015) & .623 \\
Income & & & -0.03 (0.01) & .073 \\
Sensation seeking & & & 0.02 (0.01) & .163 \\
AI literacy & & & -0.02 (0.02) & .173 \\
Digital skill literacy & & & -0.03 (0.02) & .071 \\
GenAI usage & & & 0.000 (0.015) & .994 \\
Party affiliation: Independent & & & -0.03 (0.04) & .520 \\
Party affiliation: Republican & & & -0.008 (0.032) & .810 \\
Adjusted R\textsuperscript{2} & \multicolumn{2}{>{\centering\arraybackslash}p{(\linewidth - 8\tabcolsep) * \real{0.3000} + 2\tabcolsep}}{%
0.005} & \multicolumn{2}{>{\centering\arraybackslash}p{(\linewidth - 8\tabcolsep) * \real{0.3000} + 2\tabcolsep}@{}}{%
0.017} \\
\end{longtable*}
}
\emph{Note.} Reference groups: gender, not female; race, not White; party affiliation, Democrat.
* p \textless{} .05. ** p \textless{} .01. *** p \textless{} .001.
\endgroup

\clearpage
\begingroup
\small
\setlength{\tabcolsep}{4pt}
\renewcommand{\arraystretch}{1.12}
\captionof{table}{Individual-level linear regression models predicting sharing intentions from overestimation, overplacement, and overprecision (Model 1 \emph{N} = 973; Model 2 \emph{n} = 950). Numeric predictors are standardized.}\label{tab:s13}
{ % do not increment counter
\begin{longtable*}[]{@{}
  >{\raggedright\arraybackslash}p{(\linewidth - 8\tabcolsep) * \real{0.4001}}
  >{\centering\arraybackslash}p{(\linewidth - 8\tabcolsep) * \real{0.2031}}
  >{\centering\arraybackslash}p{(\linewidth - 8\tabcolsep) * \real{0.0969}}
  >{\centering\arraybackslash}p{(\linewidth - 8\tabcolsep) * \real{0.2031}}
  >{\centering\arraybackslash}p{(\linewidth - 8\tabcolsep) * \real{0.0969}}@{}}
\toprule\noalign{}
\begin{minipage}[b]{\linewidth}\raggedright
\textbf{Predictor}
\end{minipage} & \multicolumn{2}{>{\centering\arraybackslash}p{(\linewidth - 8\tabcolsep) * \real{0.3000} + 2\tabcolsep}}{%
\begin{minipage}[b]{\linewidth}\centering
\textbf{Model 1}
\end{minipage}} & \multicolumn{2}{>{\centering\arraybackslash}p{(\linewidth - 8\tabcolsep) * \real{0.3000} + 2\tabcolsep}@{}}{%
\begin{minipage}[b]{\linewidth}\centering
\textbf{Model 2}
\end{minipage}} \\
\begin{minipage}[b]{\linewidth}\raggedright
\end{minipage} & \begin{minipage}[b]{\linewidth}\centering
\textbf{b (SE)}
\end{minipage} & \begin{minipage}[b]{\linewidth}\centering
\textbf{p}
\end{minipage} & \begin{minipage}[b]{\linewidth}\centering
\textbf{b (SE)}
\end{minipage} & \begin{minipage}[b]{\linewidth}\centering
\textbf{p}
\end{minipage} \\
\midrule\noalign{}
\endhead
\bottomrule\noalign{}
\endlastfoot
Intercept & 2.44 (0.04)*** & \textless{} .001 & 2.57 (0.09)*** & \textless{} .001 \\
Overestimation & 0.14 (0.07) & .052 & -0.02 (0.07) & .809 \\
Overplacement & -0.22 (0.07)** & .002 & -0.09 (0.06) & .177 \\
Overprecision & 0.44 (0.06)*** & \textless{} .001 & 0.33 (0.05)*** & \textless{} .001 \\
Age & & & 0.08 (0.04) & .065 \\
Gender: female & & & 0.11 (0.08) & .133 \\
Race: White & & & -0.30 (0.09)** & .001 \\
Education & & & -0.11 (0.04)** & .007 \\
Income & & & -0.07 (0.04) & .067 \\
Sensation seeking & & & 0.13 (0.04)*** & \textless{} .001 \\
AI literacy & & & -0.10 (0.05)* & .046 \\
Digital skill literacy & & & -0.03 (0.05) & .549 \\
GenAI usage & & & 0.59 (0.04)*** & \textless{} .001 \\
Party affiliation: Independent & & & -0.08 (0.12) & .522 \\
Party affiliation: Republican & & & 0.17 (0.09) & .056 \\
Adjusted R\textsuperscript{2} & \multicolumn{2}{>{\centering\arraybackslash}p{(\linewidth - 8\tabcolsep) * \real{0.3000} + 2\tabcolsep}}{%
0.088} & \multicolumn{2}{>{\centering\arraybackslash}p{(\linewidth - 8\tabcolsep) * \real{0.3000} + 2\tabcolsep}@{}}{%
0.300} \\
\end{longtable*}
}
\emph{Note.} Reference groups: gender, not female; race, not White; party affiliation, Democrat.
* p \textless{} .05. ** p \textless{} .01. *** p \textless{} .001.
\endgroup

\clearpage
\begin{landscape}
\begingroup
\footnotesize
\setlength{\tabcolsep}{4pt}
\renewcommand{\arraystretch}{1.0}
\captionof{table}{Trial-level linear mixed-effects models predicting sharing intentions. Individual overestimation and overplacement are constant across trials. Numeric predictors are standardized.}\label{tab:s14}
{ % do not increment counter
\begin{longtable*}[]{@{}
  >{\raggedright\arraybackslash}p{(\linewidth - 12\tabcolsep) * \real{0.3163}}
  >{\centering\arraybackslash}p{(\linewidth - 12\tabcolsep) * \real{0.1505}}
  >{\centering\arraybackslash}p{(\linewidth - 12\tabcolsep) * \real{0.0774}}
  >{\centering\arraybackslash}p{(\linewidth - 12\tabcolsep) * \real{0.1505}}
  >{\centering\arraybackslash}p{(\linewidth - 12\tabcolsep) * \real{0.0774}}
  >{\centering\arraybackslash}p{(\linewidth - 12\tabcolsep) * \real{0.1505}}
  >{\centering\arraybackslash}p{(\linewidth - 12\tabcolsep) * \real{0.0775}}@{}}
\toprule\noalign{}
\begin{minipage}[b]{\linewidth}\raggedright
\textbf{Predictor}
\end{minipage} & \multicolumn{2}{>{\centering\arraybackslash}p{(\linewidth - 12\tabcolsep) * \real{0.2279} + 2\tabcolsep}}{%
\begin{minipage}[b]{\linewidth}\centering
\textbf{Model 1}
\end{minipage}} & \multicolumn{2}{>{\centering\arraybackslash}p{(\linewidth - 12\tabcolsep) * \real{0.2279} + 2\tabcolsep}}{%
\begin{minipage}[b]{\linewidth}\centering
\textbf{Model 2}
\end{minipage}} & \multicolumn{2}{>{\centering\arraybackslash}p{(\linewidth - 12\tabcolsep) * \real{0.2279} + 2\tabcolsep}@{}}{%
\begin{minipage}[b]{\linewidth}\centering
\textbf{Model 3}
\end{minipage}} \\
\begin{minipage}[b]{\linewidth}\raggedright
\end{minipage} & \begin{minipage}[b]{\linewidth}\centering
\textbf{b (SE)}
\end{minipage} & \begin{minipage}[b]{\linewidth}\centering
\textbf{p}
\end{minipage} & \begin{minipage}[b]{\linewidth}\centering
\textbf{b (SE)}
\end{minipage} & \begin{minipage}[b]{\linewidth}\centering
\textbf{p}
\end{minipage} & \begin{minipage}[b]{\linewidth}\centering
\textbf{b (SE)}
\end{minipage} & \begin{minipage}[b]{\linewidth}\centering
\textbf{p}
\end{minipage} \\
\midrule\noalign{}
\endhead
\bottomrule\noalign{}
\endlastfoot
Intercept & 2.86 (0.05)*** & \textless{} .001 & 3.02 (0.10)*** & \textless{} .001 & 3.25 (0.10)*** & \textless{} .001 \\
Overestimation & 0.29 (0.07)*** & \textless{} .001 & 0.09 (0.07) & .191 & 0.05 (0.07) & .480 \\
Overplacement & -0.06 (0.07) & .346 & 0.02 (0.06) & .724 & 0.01 (0.06) & .850 \\
Overprecision & 0.048 (0.009)*** & \textless{} .001 & 0.047 (0.010)*** & \textless{} .001 & 0.37 (0.02)*** & \textless{} .001 \\
Context: health & -0.52 (0.02)*** & \textless{} .001 & -0.52 (0.02)*** & \textless{} .001 & -0.51 (0.02)*** & \textless{} .001 \\
Context: environment & -0.05 (0.02)* & .024 & -0.05 (0.02)* & .031 & -0.05 (0.02)* & .031 \\
Actual source: AI & 0.04 (0.02)* & .036 & 0.04 (0.02)* & .045 & -0.55 (0.04)*** & \textless{} .001 \\
Perceived source: AI & -0.53 (0.02)*** & \textless{} .001 & -0.52 (0.02)*** & \textless{} .001 & -0.38 (0.02)*** & \textless{} .001 \\
Age & & & 0.12 (0.04)** & .008 & 0.11 (0.04)* & .012 \\
Gender: female & & & 0.12 (0.08) & .120 & 0.11 (0.08) & .146 \\
Race: White & & & -0.35 (0.09)*** & \textless{} .001 & -0.35 (0.09)*** & \textless{} .001 \\
Education & & & -0.13 (0.04)** & .001 & -0.13 (0.04)** & .001 \\
Income & & & -0.06 (0.04) & .169 & -0.06 (0.04) & .174 \\
Sensation seeking & & & 0.14 (0.04)*** & \textless{} .001 & 0.13 (0.04)** & .001 \\
AI literacy & & & -0.10 (0.05) & .055 & -0.09 (0.05) & .065 \\
Digital skill literacy & & & -0.01 (0.05) & .746 & -0.01 (0.05) & .777 \\
GenAI usage & & & 0.60 (0.04)*** & \textless{} .001 & 0.60 (0.04)*** & \textless{} .001 \\
Party affiliation: Independent & & & -0.02 (0.12) & .836 & -0.02 (0.12) & .870 \\
Party affiliation: Republican & & & 0.20 (0.09)* & .027 & 0.20 (0.09)* & .029 \\
Overestimation \ensuremath{\times} Perceived source: AI & & & & & 0.09 (0.03)** & .004 \\
Overplacement \ensuremath{\times} Perceived source: AI & & & & & 0.009 (0.029) & .769 \\
Overprecision \ensuremath{\times} Perceived source: AI & & & & & -0.66 (0.04)*** & \textless{} .001 \\
\multicolumn{7}{@{}>{\raggedright\arraybackslash}p{(\linewidth - 12\tabcolsep) * \real{0.9998} + 12\tabcolsep}@{}}{%
\textbf{Random effects, variance (SD)}} \\
Participant intercept & \multicolumn{2}{>{\centering\arraybackslash}p{(\linewidth - 12\tabcolsep) * \real{0.2279} + 2\tabcolsep}}{%
1.70 (1.30)} & \multicolumn{2}{>{\centering\arraybackslash}p{(\linewidth - 12\tabcolsep) * \real{0.2279} + 2\tabcolsep}}{%
1.25 (1.12)} & \multicolumn{2}{>{\centering\arraybackslash}p{(\linewidth - 12\tabcolsep) * \real{0.2279} + 2\tabcolsep}@{}}{%
1.25 (1.12)} \\
Trial-order intercept & \multicolumn{2}{>{\centering\arraybackslash}p{(\linewidth - 12\tabcolsep) * \real{0.2279} + 2\tabcolsep}}{%
0.009 (0.095)} & \multicolumn{2}{>{\centering\arraybackslash}p{(\linewidth - 12\tabcolsep) * \real{0.2279} + 2\tabcolsep}}{%
0.009 (0.095)} & \multicolumn{2}{>{\centering\arraybackslash}p{(\linewidth - 12\tabcolsep) * \real{0.2279} + 2\tabcolsep}@{}}{%
0.009 (0.096)} \\
Residual & \multicolumn{2}{>{\centering\arraybackslash}p{(\linewidth - 12\tabcolsep) * \real{0.2279} + 2\tabcolsep}}{%
1.29 (1.14)} & \multicolumn{2}{>{\centering\arraybackslash}p{(\linewidth - 12\tabcolsep) * \real{0.2279} + 2\tabcolsep}}{%
1.30 (1.14)} & \multicolumn{2}{>{\centering\arraybackslash}p{(\linewidth - 12\tabcolsep) * \real{0.2279} + 2\tabcolsep}@{}}{%
1.28 (1.13)} \\
\multicolumn{7}{@{}>{\raggedright\arraybackslash}p{(\linewidth - 12\tabcolsep) * \real{0.9998} + 12\tabcolsep}@{}}{%
\textbf{Model summary}} \\
Observations & \multicolumn{2}{>{\centering\arraybackslash}p{(\linewidth - 12\tabcolsep) * \real{0.2279} + 2\tabcolsep}}{%
17,514} & \multicolumn{2}{>{\centering\arraybackslash}p{(\linewidth - 12\tabcolsep) * \real{0.2279} + 2\tabcolsep}}{%
17,100} & \multicolumn{2}{>{\centering\arraybackslash}p{(\linewidth - 12\tabcolsep) * \real{0.2279} + 2\tabcolsep}@{}}{%
17,100} \\
Participants & \multicolumn{2}{>{\centering\arraybackslash}p{(\linewidth - 12\tabcolsep) * \real{0.2279} + 2\tabcolsep}}{%
973} & \multicolumn{2}{>{\centering\arraybackslash}p{(\linewidth - 12\tabcolsep) * \real{0.2279} + 2\tabcolsep}}{%
950} & \multicolumn{2}{>{\centering\arraybackslash}p{(\linewidth - 12\tabcolsep) * \real{0.2279} + 2\tabcolsep}@{}}{%
950} \\
Marginal / conditional R\textsuperscript{2} & \multicolumn{2}{>{\centering\arraybackslash}p{(\linewidth - 12\tabcolsep) * \real{0.2279} + 2\tabcolsep}}{%
0.061 / 0.596} & \multicolumn{2}{>{\centering\arraybackslash}p{(\linewidth - 12\tabcolsep) * \real{0.2279} + 2\tabcolsep}}{%
0.204 / 0.595} & \multicolumn{2}{>{\centering\arraybackslash}p{(\linewidth - 12\tabcolsep) * \real{0.2279} + 2\tabcolsep}@{}}{%
0.210 / 0.601} \\
\end{longtable*}
}
\emph{Note.} Reference groups: gender, not female; race, not White; party affiliation, Democrat. Context, entertainment; actual and perceived source, human-made.
* p \textless{} .05. ** p \textless{} .01. *** p \textless{} .001.
\endgroup
\end{landscape}

\clearpage
\begin{landscape}
\begingroup
\small
\setlength{\tabcolsep}{4pt}
\renewcommand{\arraystretch}{1.12}
\captionof{table}{Individual-level linear regression models predicting sharing intentions (\emph{n} = 950). Numeric predictors are standardized.}\label{tab:s15}
{ % do not increment counter
\begin{longtable*}[]{@{}
  >{\raggedright\arraybackslash}p{(\linewidth - 12\tabcolsep) * \real{0.2833}}
  >{\centering\arraybackslash}p{(\linewidth - 12\tabcolsep) * \real{0.1700}}
  >{\centering\arraybackslash}p{(\linewidth - 12\tabcolsep) * \real{0.0689}}
  >{\centering\arraybackslash}p{(\linewidth - 12\tabcolsep) * \real{0.1700}}
  >{\centering\arraybackslash}p{(\linewidth - 12\tabcolsep) * \real{0.0689}}
  >{\centering\arraybackslash}p{(\linewidth - 12\tabcolsep) * \real{0.1700}}
  >{\centering\arraybackslash}p{(\linewidth - 12\tabcolsep) * \real{0.0689}}@{}}
\toprule\noalign{}
\textbf{Predictor} & \multicolumn{2}{c}{\textbf{Overestimation}} & \multicolumn{2}{c}{\textbf{Overplacement}} & \multicolumn{2}{c}{\textbf{Overprecision}} \\
\cmidrule(lr){2-3}\cmidrule(lr){4-5}\cmidrule(lr){6-7}
\begin{minipage}[b]{\linewidth}\raggedright
\end{minipage} & \begin{minipage}[b]{\linewidth}\centering
\textbf{b (SE)}
\end{minipage} & \begin{minipage}[b]{\linewidth}\centering
\textbf{p}
\end{minipage} & \begin{minipage}[b]{\linewidth}\centering
\textbf{b (SE)}
\end{minipage} & \begin{minipage}[b]{\linewidth}\centering
\textbf{p}
\end{minipage} & \begin{minipage}[b]{\linewidth}\centering
\textbf{b (SE)}
\end{minipage} & \begin{minipage}[b]{\linewidth}\centering
\textbf{p}
\end{minipage} \\
\midrule\noalign{}
\endhead
\bottomrule\noalign{}
\endlastfoot
Intercept & 2.60 (0.09)*** & \textless{} .001 & 2.60 (0.09)*** & \textless{} .001 & 2.58 (0.09)*** & \textless{} .001 \\
Overestimation & 0.12 (0.04)** & .006 & & & & \\
Overplacement & & & 0.10 (0.04)* & .016 & & \\
Overprecision & & & & & 0.27 (0.04)*** & \textless{} .001 \\
Age & 0.13 (0.04)** & .005 & 0.13 (0.05)** & .006 & 0.07 (0.04) & .098 \\
Gender: female & 0.11 (0.08) & .161 & 0.11 (0.08) & .159 & 0.11 (0.08) & .133 \\
Race: White & -0.35 (0.09)*** & \textless{} .001 & -0.36 (0.09)*** & \textless{} .001 & -0.31 (0.09)*** & \textless{} .001 \\
Education & -0.14 (0.04)** & .001 & -0.14 (0.04)** & .001 & -0.12 (0.04)** & .004 \\
Income & -0.06 (0.04) & .140 & -0.06 (0.04) & .155 & -0.07 (0.04) & .079 \\
Sensation seeking & 0.14 (0.04)*** & \textless{} .001 & 0.15 (0.04)*** & \textless{} .001 & 0.14 (0.04)*** & \textless{} .001 \\
AI literacy & -0.10 (0.05)* & .037 & -0.07 (0.05) & .119 & -0.11 (0.05)* & .012 \\
Digital skill literacy & -0.02 (0.05) & .691 & -0.03 (0.05) & .571 & -0.03 (0.05) & .528 \\
GenAI usage & 0.61 (0.04)*** & \textless{} .001 & 0.61 (0.04)*** & \textless{} .001 & 0.58 (0.04)*** & \textless{} .001 \\
Party affiliation: Independent & -0.03 (0.12) & .820 & -0.02 (0.12) & .861 & -0.06 (0.12) & .596 \\
Party affiliation: Republican & 0.20 (0.09)* & .028 & 0.21 (0.09)* & .020 & 0.17 (0.09) & .061 \\
Adjusted R\textsuperscript{2} & \multicolumn{2}{>{\centering\arraybackslash}p{(\linewidth - 12\tabcolsep) * \real{0.2389} + 2\tabcolsep}}{%
0.273} & \multicolumn{2}{>{\centering\arraybackslash}p{(\linewidth - 12\tabcolsep) * \real{0.2389} + 2\tabcolsep}}{%
0.271} & \multicolumn{2}{>{\centering\arraybackslash}p{(\linewidth - 12\tabcolsep) * \real{0.2389} + 2\tabcolsep}@{}}{%
0.299} \\
\end{longtable*}
}
\emph{Note.} Reference groups: gender, not female; race, not White; party affiliation, Democrat.
* p \textless{} .05. ** p \textless{} .01. *** p \textless{} .001.
\endgroup
\end{landscape}

\clearpage
\begin{landscape}
\begingroup
\footnotesize
\setlength{\tabcolsep}{4pt}
\renewcommand{\arraystretch}{1.0}
\captionof{table}{Trial-level linear mixed-effects models predicting sharing intentions. Individual overestimation and overplacement are constant across trials. Numeric predictors are standardized.}\label{tab:s16}
{ % do not increment counter
\begin{longtable*}[]{@{}
  >{\raggedright\arraybackslash}p{(\linewidth - 12\tabcolsep) * \real{0.2833}}
  >{\centering\arraybackslash}p{(\linewidth - 12\tabcolsep) * \real{0.1700}}
  >{\centering\arraybackslash}p{(\linewidth - 12\tabcolsep) * \real{0.0689}}
  >{\centering\arraybackslash}p{(\linewidth - 12\tabcolsep) * \real{0.1700}}
  >{\centering\arraybackslash}p{(\linewidth - 12\tabcolsep) * \real{0.0689}}
  >{\centering\arraybackslash}p{(\linewidth - 12\tabcolsep) * \real{0.1700}}
  >{\centering\arraybackslash}p{(\linewidth - 12\tabcolsep) * \real{0.0689}}@{}}
\toprule\noalign{}
\textbf{Predictor} & \multicolumn{2}{c}{\textbf{Overestimation}} & \multicolumn{2}{c}{\textbf{Overplacement}} & \multicolumn{2}{c}{\textbf{Overprecision}} \\
\cmidrule(lr){2-3}\cmidrule(lr){4-5}\cmidrule(lr){6-7}
\begin{minipage}[b]{\linewidth}\raggedright
\end{minipage} & \begin{minipage}[b]{\linewidth}\centering
\textbf{b (SE)}
\end{minipage} & \begin{minipage}[b]{\linewidth}\centering
\textbf{p}
\end{minipage} & \begin{minipage}[b]{\linewidth}\centering
\textbf{b (SE)}
\end{minipage} & \begin{minipage}[b]{\linewidth}\centering
\textbf{p}
\end{minipage} & \begin{minipage}[b]{\linewidth}\centering
\textbf{b (SE)}
\end{minipage} & \begin{minipage}[b]{\linewidth}\centering
\textbf{p}
\end{minipage} \\
\midrule\noalign{}
\endhead
\bottomrule\noalign{}
\endlastfoot
Intercept & 3.02 (0.10)*** & \textless{} .001 & 3.02 (0.10)*** & \textless{} .001 & 3.02 (0.10)*** & \textless{} .001 \\
Overestimation & 0.12 (0.04)** & .007 & & & & \\
Overplacement & & & 0.09 (0.04)* & .020 & & \\
Overprecision & & & & & 0.048 (0.009)*** & \textless{} .001 \\
Context: health & -0.52 (0.02)*** & \textless{} .001 & -0.52 (0.02)*** & \textless{} .001 & -0.52 (0.02)*** & \textless{} .001 \\
Context: environment & -0.04 (0.02) & .058 & -0.04 (0.02) & .058 & -0.05 (0.02)* & .031 \\
Actual source: AI & 0.04 (0.02)* & .024 & 0.04 (0.02)* & .024 & 0.04 (0.02)* & .045 \\
Perceived source: AI & -0.52 (0.02)*** & \textless{} .001 & -0.52 (0.02)*** & \textless{} .001 & -0.52 (0.02)*** & \textless{} .001 \\
Age & 0.12 (0.04)** & .006 & 0.12 (0.04)** & .006 & 0.14 (0.04)** & .001 \\
Gender: female & 0.12 (0.08) & .122 & 0.12 (0.08) & .121 & 0.12 (0.08) & .125 \\
Race: White & -0.35 (0.09)*** & \textless{} .001 & -0.36 (0.09)*** & \textless{} .001 & -0.36 (0.09)*** & \textless{} .001 \\
Education & -0.13 (0.04)** & .001 & -0.13 (0.04)** & .001 & -0.13 (0.04)** & .001 \\
Income & -0.06 (0.04) & .173 & -0.05 (0.04) & .189 & -0.05 (0.04) & .178 \\
Sensation seeking & 0.13 (0.04)*** & \textless{} .001 & 0.14 (0.04)*** & \textless{} .001 & 0.14 (0.04)*** & \textless{} .001 \\
AI literacy & -0.10 (0.05)* & .044 & -0.07 (0.05) & .137 & -0.05 (0.04) & .286 \\
Digital skill literacy & -0.01 (0.05) & .783 & -0.02 (0.05) & .661 & -0.02 (0.05) & .712 \\
GenAI usage & 0.60 (0.04)*** & \textless{} .001 & 0.61 (0.04)*** & \textless{} .001 & 0.62 (0.04)*** & \textless{} .001 \\
Party affiliation: Independent & -0.02 (0.12) & .844 & -0.02 (0.12) & .883 & -0.03 (0.12) & .832 \\
Party affiliation: Republican & 0.20 (0.09)* & .026 & 0.21 (0.09)* & .019 & 0.22 (0.09)* & .013 \\
\multicolumn{7}{@{}>{\raggedright\arraybackslash}p{(\linewidth - 12\tabcolsep) * \real{1.0000} + 12\tabcolsep}@{}}{%
\textbf{Random effects, variance (SD)}} \\
Participant intercept & \multicolumn{2}{>{\centering\arraybackslash}p{(\linewidth - 12\tabcolsep) * \real{0.2389} + 2\tabcolsep}}{%
1.26 (1.12)} & \multicolumn{2}{>{\centering\arraybackslash}p{(\linewidth - 12\tabcolsep) * \real{0.2389} + 2\tabcolsep}}{%
1.26 (1.12)} & \multicolumn{2}{>{\centering\arraybackslash}p{(\linewidth - 12\tabcolsep) * \real{0.2389} + 2\tabcolsep}@{}}{%
1.26 (1.12)} \\
Trial-order intercept & \multicolumn{2}{>{\centering\arraybackslash}p{(\linewidth - 12\tabcolsep) * \real{0.2389} + 2\tabcolsep}}{%
0.009 (0.094)} & \multicolumn{2}{>{\centering\arraybackslash}p{(\linewidth - 12\tabcolsep) * \real{0.2389} + 2\tabcolsep}}{%
0.009 (0.094)} & \multicolumn{2}{>{\centering\arraybackslash}p{(\linewidth - 12\tabcolsep) * \real{0.2389} + 2\tabcolsep}@{}}{%
0.009 (0.095)} \\
Residual & \multicolumn{2}{>{\centering\arraybackslash}p{(\linewidth - 12\tabcolsep) * \real{0.2389} + 2\tabcolsep}}{%
1.30 (1.14)} & \multicolumn{2}{>{\centering\arraybackslash}p{(\linewidth - 12\tabcolsep) * \real{0.2389} + 2\tabcolsep}}{%
1.30 (1.14)} & \multicolumn{2}{>{\centering\arraybackslash}p{(\linewidth - 12\tabcolsep) * \real{0.2389} + 2\tabcolsep}@{}}{%
1.30 (1.14)} \\
\multicolumn{7}{@{}>{\raggedright\arraybackslash}p{(\linewidth - 12\tabcolsep) * \real{1.0000} + 12\tabcolsep}@{}}{%
\textbf{Model summary}} \\
Observations & \multicolumn{2}{>{\centering\arraybackslash}p{(\linewidth - 12\tabcolsep) * \real{0.2389} + 2\tabcolsep}}{%
17,100} & \multicolumn{2}{>{\centering\arraybackslash}p{(\linewidth - 12\tabcolsep) * \real{0.2389} + 2\tabcolsep}}{%
17,100} & \multicolumn{2}{>{\centering\arraybackslash}p{(\linewidth - 12\tabcolsep) * \real{0.2389} + 2\tabcolsep}@{}}{%
17,100} \\
Participants & \multicolumn{2}{>{\centering\arraybackslash}p{(\linewidth - 12\tabcolsep) * \real{0.2389} + 2\tabcolsep}}{%
950} & \multicolumn{2}{>{\centering\arraybackslash}p{(\linewidth - 12\tabcolsep) * \real{0.2389} + 2\tabcolsep}}{%
950} & \multicolumn{2}{>{\centering\arraybackslash}p{(\linewidth - 12\tabcolsep) * \real{0.2389} + 2\tabcolsep}@{}}{%
950} \\
Marginal / conditional R\textsuperscript{2} & \multicolumn{2}{>{\centering\arraybackslash}p{(\linewidth - 12\tabcolsep) * \real{0.2389} + 2\tabcolsep}}{%
0.202 / 0.595} & \multicolumn{2}{>{\centering\arraybackslash}p{(\linewidth - 12\tabcolsep) * \real{0.2389} + 2\tabcolsep}}{%
0.202 / 0.595} & \multicolumn{2}{>{\centering\arraybackslash}p{(\linewidth - 12\tabcolsep) * \real{0.2389} + 2\tabcolsep}@{}}{%
0.200 / 0.595} \\
\end{longtable*}
}
\emph{Note.} Reference groups: gender, not female; race, not White; party affiliation, Democrat. Context, entertainment; actual and perceived source, human-made.
* p \textless{} .05. ** p \textless{} .01. *** p \textless{} .001.
\endgroup
\end{landscape}

\clearpage
\begin{landscape}
\begingroup
\small
\setlength{\tabcolsep}{4pt}
\renewcommand{\arraystretch}{1.12}
\captionof{table}{\textbf{\hfill\break
} Individual characteristics associated with overestimation, overplacement, and overprecision (\emph{n} = 950). Numeric predictors and outcomes are standardized. Significance tests use Benjamini-Hochberg-adjusted p-values (q) across 33 comparisons.}\label{tab:s17}
{ % do not increment counter
\begin{longtable*}[]{@{}
  >{\raggedright\arraybackslash}p{(\linewidth - 12\tabcolsep) * \real{0.2619}}
  >{\centering\arraybackslash}p{(\linewidth - 12\tabcolsep) * \real{0.1329}}
  >{\centering\arraybackslash}p{(\linewidth - 12\tabcolsep) * \real{0.0789}}
  >{\centering\arraybackslash}p{(\linewidth - 12\tabcolsep) * \real{0.1711}}
  >{\centering\arraybackslash}p{(\linewidth - 12\tabcolsep) * \real{0.0789}}
  >{\centering\arraybackslash}p{(\linewidth - 12\tabcolsep) * \real{0.1843}}
  >{\centering\arraybackslash}p{(\linewidth - 12\tabcolsep) * \real{0.0921}}@{}}
\toprule\noalign{}
\begin{minipage}[b]{\linewidth}\raggedright
\textbf{Predictor}
\end{minipage} & \multicolumn{2}{>{\centering\arraybackslash}p{(\linewidth - 12\tabcolsep) * \real{0.2118} + 2\tabcolsep}}{%
\begin{minipage}[b]{\linewidth}\centering
\textbf{Overestimation}
\end{minipage}} & \multicolumn{2}{>{\centering\arraybackslash}p{(\linewidth - 12\tabcolsep) * \real{0.2500} + 2\tabcolsep}}{%
\begin{minipage}[b]{\linewidth}\centering
\textbf{Overplacement}
\end{minipage}} & \multicolumn{2}{>{\centering\arraybackslash}p{(\linewidth - 12\tabcolsep) * \real{0.2763} + 2\tabcolsep}@{}}{%
\begin{minipage}[b]{\linewidth}\centering
\textbf{Overprecision}
\end{minipage}} \\
\begin{minipage}[b]{\linewidth}\raggedright
\end{minipage} & \begin{minipage}[b]{\linewidth}\centering
\textbf{b (SE)}
\end{minipage} & \begin{minipage}[b]{\linewidth}\centering
\textbf{q}
\end{minipage} & \begin{minipage}[b]{\linewidth}\centering
\textbf{b (SE)}
\end{minipage} & \begin{minipage}[b]{\linewidth}\centering
\textbf{q}
\end{minipage} & \begin{minipage}[b]{\linewidth}\centering
\textbf{b (SE)}
\end{minipage} & \begin{minipage}[b]{\linewidth}\centering
\textbf{q}
\end{minipage} \\
\midrule\noalign{}
\endhead
\bottomrule\noalign{}
\endlastfoot
Intercept & -0.005 (0.068) & & -0.02 (0.07) & & 0.08 (0.07) & \\
Age & 0.21 (0.03)*** & \textless{} .001 & 0.27 (0.04)*** & \textless{} .001 & 0.29 (0.03)*** & \textless{} .001 \\
Gender: female & -0.01 (0.06) & .919 & -0.02 (0.06) & .913 & -0.03 (0.06) & .826 \\
Race: White & -0.07 (0.07) & .500 & -0.01 (0.08) & .919 & -0.19 (0.07)* & .026 \\
Education & 0.005 (0.030) & .919 & 0.01 (0.03) & .826 & -0.07 (0.03) & .099 \\
Income & 0.01 (0.03) & .826 & -0.005 (0.033) & .919 & 0.04 (0.03) & .323 \\
Sensation seeking & 0.07 (0.03)* & .034 & -0.01 (0.03) & .826 & 0.04 (0.03) & .349 \\
AI literacy & 0.48 (0.03)*** & \textless{} .001 & 0.27 (0.04)*** & \textless{} .001 & 0.26 (0.04)*** & \textless{} .001 \\
Digital skill literacy & -0.03 (0.03) & .524 & 0.04 (0.04) & .500 & 0.02 (0.04) & .781 \\
GenAI usage & 0.16 (0.03)*** & \textless{} .001 & 0.11 (0.03)** & .002 & 0.15 (0.03)*** & \textless{} .001 \\
Party affiliation: Independent & 0.006 (0.087) & .944 & -0.05 (0.10) & .818 & 0.13 (0.09) & .316 \\
Party affiliation: Republican & 0.22 (0.07)** & .003 & 0.15 (0.07) & .077 & 0.22 (0.07)** & .006 \\
Adjusted R\textsuperscript{2} & \multicolumn{2}{>{\centering\arraybackslash}p{(\linewidth - 12\tabcolsep) * \real{0.2118} + 2\tabcolsep}}{%
0.278} & \multicolumn{2}{>{\centering\arraybackslash}p{(\linewidth - 12\tabcolsep) * \real{0.2500} + 2\tabcolsep}}{%
0.136} & \multicolumn{2}{>{\centering\arraybackslash}p{(\linewidth - 12\tabcolsep) * \real{0.2763} + 2\tabcolsep}@{}}{%
0.158} \\
\end{longtable*}
}
\emph{Note.} Reference groups: gender, not female; race, not White; party affiliation, Democrat.
* q \textless{} .05. ** q \textless{} .01. *** q \textless{} .001
\endgroup
\end{landscape}

\end{document}